\documentclass[12pt,preprint]{aastex}

\def\AaA{{\em Astr.~Astrophys.}}

\def\AIPC{{\em Amer.~Inst.~Phys.~Conf.}}
\def\AJ{{\em Astr.~J.}}

\def\ApJ{{\em Astrophys.~J.}}

\def\ApJS{{\em Astrophys.~J.~Suppl.}}

\def\AZ{{\em Astronom. Zhurnal}}
\def\BAAS{{\em Bull.~Amer.~Astron.~Soc.}}
\def\GeCoA{{\em Geo.~Cosm.~Acta}}
\def\MN{{\em Mon.~Not.~R.~astr.~Soc.}}
\def\Nat{{\em Nature}}
\def\NA{{\em New~Astr.}}

\def\PR{{\em Phys.~Reports}}

\def\SAI{{\em Soc.~Astr.~Ital.~Mem.}}

\def\etal{{et al.\thinspace}}

\def\spose#1{\hbox to 0pt{#1\hss}}

\let\approxlt=\lesssim

\def\multleft#1{\hbox to size{\vbox {\halign {\lft{##}\cr #1}}\hfill}\par}
\def\multright#1{\hbox to size{\vbox {\halign {\rt{##}\cr #1}}\hfill}\par}

\def\boxit#1{\vbox{\hrule\hbox{\vrule\kern3pt\vbox{\kern3pt
          #1 \kern3pt}\kern3pt\vrule}\hrule}}

\def\cm{{\rm\thinspace cm}}

\def\erg{{\rm\thinspace erg}}
\def\eV{{\rm\thinspace eV}}

\def\K{{\rm\thinspace K}}
\def\keV{{\rm\thinspace keV}}
\def\km{{\rm\thinspace km}}

\def\Mpc{{\rm\thinspace Mpc}}
\def\Msun{\hbox{$\rm\thinspace M_{\odot}$}}

\def\ph{{\rm\thinspace ph}}
\def\s{{\rm\thinspace s}}
\def\ks{{\rm\thinspace ks}}

\def\Hz{{\rm\thinspace Hz}}
\def\chisq{\hbox{$\chi^2$}}

\def\cts{{\rm\thinspace cts}}
\def\bin{{\rm\thinspace bin}}

\def\ergcmps{\hbox{$\erg\cm\s^{-1}\,$}}
\def\ergpcmps{\hbox{$\erg\cm^{-1}\s^{-1}\,$}}
\def\ergpcmsqps{\hbox{$\erg\cm^{-2}\s^{-1}\,$}}

\def\ergps{\hbox{$\erg\s^{-1}\,$}}

\def\kmps{\hbox{$\km\s^{-1}\,$}}

\def\pcmsq{\hbox{$\cm^{-2}\,$}}

\def\phpcmsqps{\hbox{$\ph\cm^{-2}\s^{-1}\,$}}

\def\kmpspMpc{\hbox{$\kmps\Mpc^{-1}$}}

\def\ctsps{\hbox{$\cts\s^{-1}$}}
\def\ctspb{\hbox{$\cts\bin^{-1}$}}

\makeatletter

\let\@internalcite\cite
\def\cite{\@ifstar{\citey}{\citefull}}
\def\citefull{\def\astroncite##1##2{##1\ ##2}\@internalcite}
\def\citey{\def\astroncite##1##2{##1\ (##2)}\@internalcite}
\def\citeyear{\def\astroncite##1##2{##2}\@internalcite}
\def\citename{\def\astroncite##1##2{##1}\@internalcite}
\def\@citex[#1]#2{\if@filesw\immediate\write\@auxout{\string\citation{#2}}\fi
  \def\@citea{}\@cite{\@for\@citeb:=#2\do
    {\@citea\def\@citea{; }\@ifundefined
       {b@\@citeb}{{\bf ??}\@warning
       {Citation `\@citeb' on page \thepage \space undefined}}%
{\csname b@\@citeb\endcsname}}}{#1}}
 
\def\@cite#1#2{#1\if@tempswa #2\fi} 
\def\@biblabel#1{}
 
\def\astroncite#1#2{#1\ #2}
\makeatother

\begin{document}

\title{An X-ray Spectral Analysis of the Central Regions of NGC 4593}

\author{Laura~W.~Brenneman\altaffilmark{1}, 
Christopher~S.~Reynolds\altaffilmark{1},
J\"{o}rn~Wilms\altaffilmark{2}, and
Mary~Elizabeth~Kaiser\altaffilmark{3}}

\altaffiltext{1}{Dept. of Astronomy, University of Maryland, College
Park, College Park MD~20742}
\altaffiltext{2}{c/o Dr. Karl Remeis-Sternwarte, University of
  Erlangen-Nuremberg, Sternwartstr. 7, 96049 Bamberg, Germany}
\altaffiltext{3}{Dept. of Physics and Astronomy, Johns Hopkins University, 
3400 North Charles Street, Baltimore MD~21212}

\begin{abstract}
We present a detailed analysis of {\it XMM-Newton} EPIC-pn data for
the Seyfert-1 galaxy NGC~4593.  We discuss the X-ray spectral
properties of this source as well as its variations with time.  The
$0.5-10 \keV$ spectrum shows significant complexity beyond a simple
power-law form, with clear evidence existing for a ``soft excess'' as
well as absorption by highly ionized plasma (a warm absorber) within
the central engine of this active galactic nucleus.  We show that the
soft excess is best described as originating from thermal
Comptonization by plasma that is appreciably cooler than the primary
X-ray emitting plasma; we find that the form of the soft excess cannot
be reproduced adequately by reflection from an ionized accretion disk.
The only measurable deviation from the power-law continuum in the hard
spectrum comes from the presence of cold and ionized fluorescent iron
K$\alpha$ emission lines at $6.4$ and $6.97 \keV$, respectively.
While constraints on the ionized iron line are weak, the cold line is
found to be narrow at CCD-resolution with a flux that does not track
the temporal changes in the underlying continuum, implying an origin
in the outer radii of the accretion disk or the putative molecular
torus of Seyfert unification schemes.  The X-ray continuum itself
varies on all accessible time scales.  We detect a $\sim 230 \s$
time-lag between soft and hard EPIC-pn bands that, if interpreted as
scattering timescales within a Comptonizing disk corona, can be used
to constrain the physical size of the primary X-ray source to a
characteristic length scale of $\sim 2 r_{\rm g}$.  Taken together,
the small implied coronal size and the large implied iron line
emitting region indicate a departure from the current picture of a
``typical'' AGN geometry.   
\end{abstract}


\section{Introduction}
\label{sec:intro}

It is thought that supermassive black holes are the central engines of
active galactic nuclei, and may inhabit the cores of most galaxies
(Salpeter 1964, Zel'Dovich 1965, Lynden-Bell 1969).  Accretion disks
around such objects provide some of the best natural laboratories for
examining the effects of an extreme gravity environment on infalling
material.  With the enhanced throughput and spectral resolution of
modern X-ray telescopes such as {\it Chandra}, {\it XMM-Newton} and
{\it Suzaku}, it has now become possible to observe these effects with
the precision necessary to make robust measurements of the innermost
properties and overall structures of accretion disks.  Evidence for
relativistic emission line broadening, spectral and temporal
variability over a wide range of time scales, and photoionized (warm)
absorption from outflowing highly-ionized gas along the line of sight
are only some of the interesting spectral detections being made in
many active galaxies.

In this paper, we present an X-ray spectral analysis of the central
regions of the canonical Seyfert-1 galaxy, NGC~4593.  Galaxies of this
type are of particular interest for the study of accretion onto
supermassive black holes since they are typically oriented such that
we can view the accreting black hole free of substantial obscuration
or absorption from surrounding circumnuclear material.  Furthermore,
we believe that significant amounts of the continuum radiation seen in
radio-quiet Sy-1 galaxies originate in the disk rather than, for example, a
relativistically beamed jet (as is the case for BL-Lac objects).
As such, we are better positioned to study the inner parts of the 
accretion disk in Sy-1 galaxies than in other systems.

NGC~4593 is a spiral galaxy with a central bar classified as Hubble
type SBb.  At a redshift of $z = 0.009$, it lies at a proper distance
of $37.9 \Mpc$ toward the constellation Virgo.  This corresponds to an
angular size distance of $37.6 \Mpc$, and a luminosity distance of
$38.3 \Mpc$ using $H_0 = 71 \kmpspMpc$, $\Omega_{\rm m} = 0.27$, and
$\Omega_{\Lambda} = 0.73$.\footnote{These values have been
obtained from Ned Wright's Cosmology Calculator web page:
http://www.astro.ucla.edu/~wright/CosmoCalc.html.}  The galaxy has
an apparent visual magnitude of $11.67$ and an approximate angular
diameter of $3.9\times 2.9$ arcmin.  As already noted, it hosts an
active galactic nucleus (AGN) of a Sy-1 type (Lewis, MacAlpine \&
Koski 1978).  Previous studies of the source with {\it EXOSAT}
demonstrated a soft excess (Pounds \& Turner 1988), and {\it BeppoSAX}
data confirm a broad absorption dip of $15\%$ below $1 \keV$ which may
be attributable to the presence of a warm absorber along the line of
sight (Kaastra \& Steenbrugge 2001).  {\it ASCA} spectra display a
slightly broadened cold iron line at $6.4 \keV$, in addition to
evidence for a warm absorber within the system (Nandra \etal 1997;
Reynolds 1997).  The source also displays significant variability in
flux.  Between two {\it ASCA} observations $3.5$ years apart, the
$2-10 \keV$ flux of this source increased by $\sim 25\%$, though no
significant variability of the iron line was detected between the two
pointings (Weaver \etal 2001).  Within the {\it ASCA} observation,
Reynolds (1997) witnessed a count rate decrease of $\sim 60\%$ in $10
\ks$, with smaller flares and dips throughout the data set.  In terms
of the overall properties of the system, Reynolds (1997) calculated a
luminosity of $L_{\rm X} = 8.53 \times 10^{42} \ergps$ ($2-10 \keV$)
with {\it ASCA}, and more recently, McKernan \etal (2003) used {\it
Chandra} to derive a luminosity $L_{\rm X} = 5.37 \times 10^{42}
\ergps$ ($2-10 \keV$).  It should be noted that this change in $L_{\rm
X}$ is larger than the flux calibration uncertainty of the
observation, so this does appear to be a robust finding.

In this paper, we present a $76\ks$ exposure of NGC~4593 with the {\it
XMM-Newton}/EPIC-pn instrument from 2002 June 23/24.  We discuss our
observations of the time-averaged $2-10 \keV$ spectrum and cold iron
line in more detail in Reynolds \etal (2004; hereafter R04).  Here, we
will examine the spectral and temporal variability of NGC~4593, as
well as the soft X-ray spectrum.  \S2 discusses the method and
software employed for the reduction of the data referred to in this
work.  Spectral fitting analysis is detailed for the EPIC-pn
instrument in \S3, and variability studies are examined in \S4.  We
present our discussion in \S5, including a comparison of our results
with those published in earlier studies of NGC~4593 and a discussion
of the disparate spatial scales implied for the X-ray emitting region
and the iron fluorescence region.  We also draw our conclusions in
\S5.

\section{Data Reduction}
\label{sec:reduction}

In this paper, we use data taken with the European Photon Imaging
Camera pn (EPIC-pn) camera.  The data were obtained during revolution
465 of {\it XMM-Newton}, during which the pn was operated in its
small-window mode to prevent photon pile up, using the medium filter
to avoid optical light contamination.  The EPIC MOS-1 camera took data
in the fast uncompressed timing mode, and the MOS-2 camera operated in
prime partial W2 imaging mode.  Although the MOS results will not be
discussed further here, they mirrored the EPIC-pn data within the
expected errors of calibration effects.  The average EPIC-pn count
rate for this source was $29.78 \ctsps$.  

Data were also collected during the observation by the Reflection
Grating Spectrometer (RGS).  These data were examined but were deemed
to be unhelpful for the analysis presented in this paper.  Firstly,
the RGS spectra have insufficient signal-to-noise to allow a detailed
study of the soft X-ray emission and absorption lines from the warm
absorber in this source.  They also contain insufficient counts to
allow variability analysis.  Finally, our examination of the RGS
spectra have demonstrated that problems with the {\it broad-band}
effective area calibration of the RGS (which is strongly affected by a
build-up of carbon on the CCDs) compromises any RGS-based results on
the shape of the soft excess.  For a detailed discussion of these
calibration uncertainties, see Kirsch \etal (2005).\footnote{See also
http://xmm.vilspa.esa.es/external/xmm\_sw\_cal/calib/index.shtml for a
thorough discussion of these cross-calibration issues.}  For these
reasons, will shall not consider RGS data further.
 
The pipeline data were reprocessed using the Science Analysis Software
and the corresponding calibration files, version 6.5.0.  From these,
we rebuilt the calibration index file using {\tt cifbuild}.  For the
EPIC-pn data the event files were mildly edited in spectral coverage
to observe the region from $0.2-15 \keV$, and bad pixels and cosmic
ray spikes were removed via narrow time filtering using the {\tt
evselect} task within the SAS.  No background flares were detected
during the observation.  Extraction of spectra followed the procedure
used by Wilms \etal (2001), in which source and background spectra
were generated using the {\tt xmmselect} task.  Response matrices and
ancillary response files were created using {\tt rmfgen} and {\tt
arfgen}, and the data were then grouped using the {\tt grppha} task
with a binning factor of $25 \ctspb$.  Binning is required in order to
get sufficient counts per bin to make $\chisq$ spectral fitting a
valid statistical process.  Spectral modeling and analysis from
$0.5-10 \keV$ was performed using the {\sc xspec} package version
11.3.2. (Arnaud 1996).  Timing studies were performed using various
routines in the {\sc xronos} package (Stella \& Angelini 1991).

We have used the SAS {\tt epatplot} task to compute the fraction of single,
double, triple and quadruple events as a function of energy and compared
these fractions to their nominal values as measured from weak source
observations. For sources that are affected by pile up, these fractions
deviate from the nominal values due to the higher probability of wrong
pattern classification. No significant deviation from the nominal single
and double distributions was found, indicating that our EPIC-pn
observation of NGC~4593 is not affected by pile up.

\section{Spectral Analysis}
\label{sec:spectra}

As shown in R04, the 2--10\,keV EPIC-pn spectrum of NGC~4593 is very
well described by a power-law with the exception of two emission
features identified as the fluorescent K$\alpha$ emission line of cold
iron at 6.4\,keV and the Ly$\alpha$ recombination line of
hydrogen-like iron at 6.97\,keV (Fig.~1).  The cold iron line likely
arises from fluorescence on the surface layers of the outer accretion
disk, optically-thick optical broad emission line clouds, or the
putative ``molecular torus'' in response to hard X-ray irradiation
from a hot corona in the inner accretion disk (Basko 1978; Guilbert \&
Rees 1988; Lightman \& White 1988; George \& Fabian 1991; Matt \etal
1991).  The ionized line, by contrast, could be formed either by
ionized disk irradiation or by radiative recombination in
highly-ionized outflowing material above the plane of the disk.

\begin{figure}
\centerline{ \includegraphics[scale=.6,angle=270]{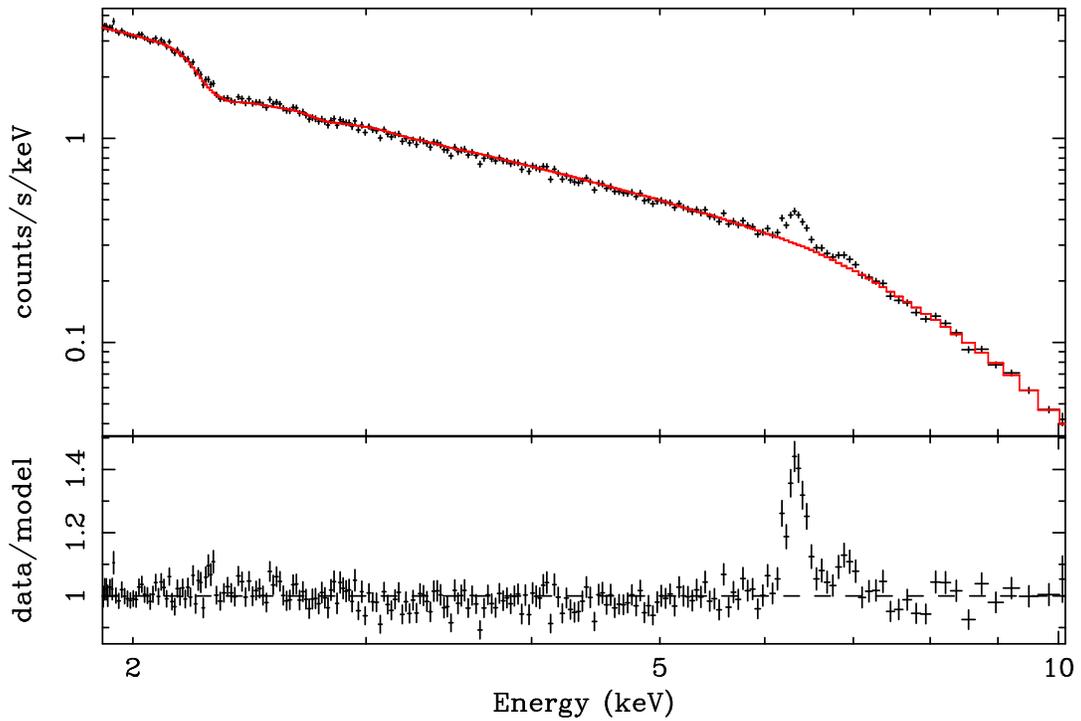} }
\caption{The $2-10 \keV$ spectrum of NGC~4593 fit with a simple
  photoabsorbed power-law ({\tt phabs po}).  Note the residual
  iron features at $6.4$ and $6.97 \keV$.  $\chi^2/{\rm
    dof}=1879/1450 (1.30)$.}
\end{figure}

However, extrapolating below $2 \keV$ reveals a clear soft excess as
well as the effects of the warm absorber (Fig.~2; the ``notch'' in the
range 0.7--1.0\,keV signals the presence of the warm absorber).  Since
the basic origin of such soft excesses are still the subject of much
debate, it is interesting to characterize the soft excess in as much
detail as possible.  This necessitates, however, a detailed modeling
of the warm absorber.

We model the effect of the warm absorber using the {\sc xstar}
photoionization code (version 2.1kn3; originally developed by Kallman
\& Krolik 1995).  In detail, we construct grids of models describing
absorption by photoionized slabs of matter with column density $N_H$
and ionization parameter $\xi$, \begin{equation}
\xi=\frac{L_{\rm i}}{n_{\rm e}r^2},
\end{equation}
where $L_{\rm i}$ is the luminosity above the hydrogen Lyman limit,
$n_{\rm e}$ is the electron number density of the plasma and $r$ is
the distance from the (point) source is ionizing luminosity.  We have
constructed $20\times 20$ grids of models uniformly sampling the
$(\log\,N_{\rm H},\log\xi)$ plane in the range $N_{\rm
H}=10^{20}\rightarrow 10^{24}\pcmsq$ and $\xi=1\rightarrow
10^4\ergpcmps$.  While these were made to be multiplicative absorption
models and hence can be applied to any emission spectrum, the
ionization balance was solved assuming a power-law ionizing spectrum
with a photon index of $\Gamma=2$, cut off at an energy $E_{\rm c}=20
\keV$.  This is a good approximation of the typical AGN continuum.

\begin{figure}
\centerline{
\includegraphics[scale=.6,angle=270]{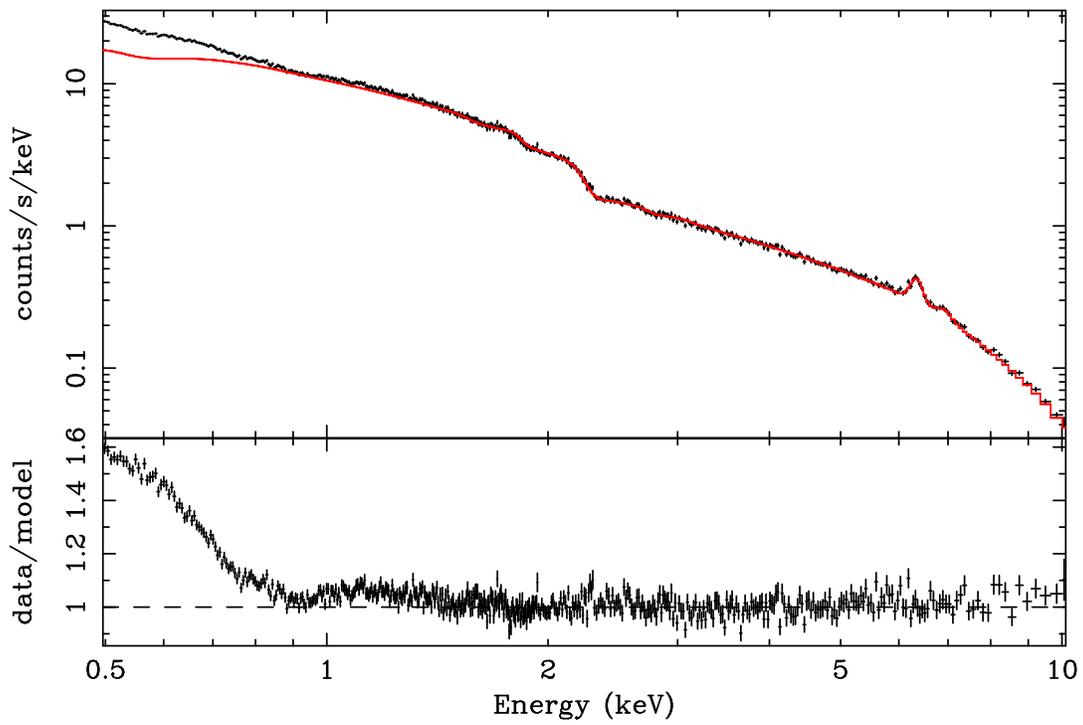}
}
\caption{The $0.5-10.0 \keV$ spectrum of NGC~4593 fit with the
  photoabsorbed power-law from Fig.~1 and two Gaussians to model the
  iron lines.  Note the clear evidence for a soft excess, possibly
  complicated by absorption features from a ``warm absorber'' within
  NGC 4593.  For this fit, $\chi^2/{\rm dof}=11423/1736 (6.58)$.}
\end{figure}

As a first, purely phenomenalogical attempt to describe the soft
excess component, we have employed a single temperature thermal
bremsstrahlung emission model ({\tt zbremss}).  To adequately describe
the spectrum, we were required to include two warm absorbing zones as
well as an iron-L$_3$ edge from iron atoms in line-of-sight dust.  A
natural interpretation is that the WA has distinct components at
physically different distances from the central engine.  We note that
attempting to add a third zone into the model results in no
statistical improvement in fit.  Including this absorption structure
provides a good description of the data ($\chi^2=1813/1745$; Fig.~3
and Table~1).  We shall refer to this model as Model-1.  The best fit
value for the temperature characterizing this soft excess component is
$kT \approx 0.21^{+0.003}_{-0.005} \keV$ and it has a 0.5--10\,keV
flux of $2.01^{+0.33}_{-0.11} \times 10^{-14} \ergpcmsqps$.

\begin{figure}
\includegraphics[scale=.5,angle=270]{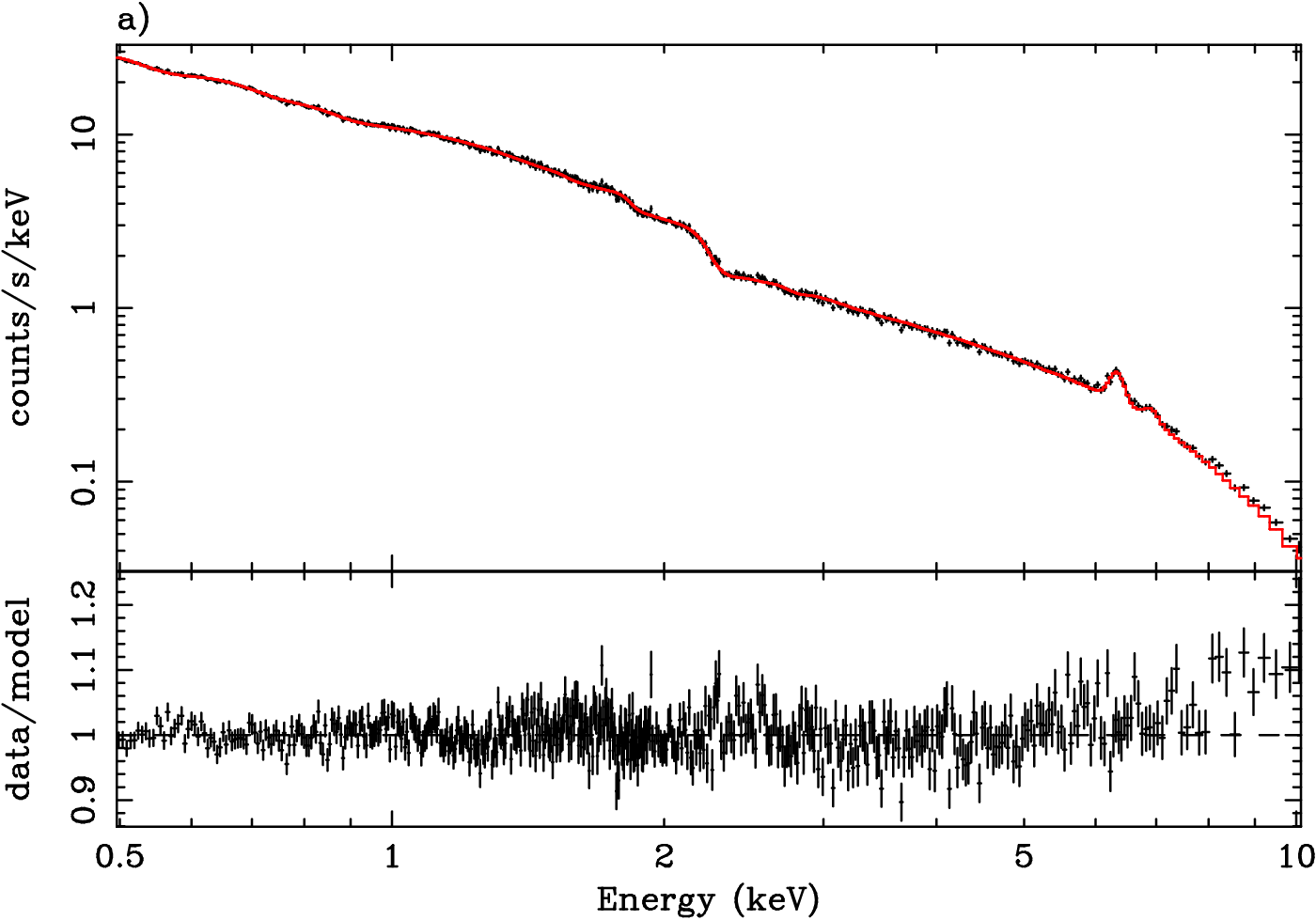}\\
\includegraphics[scale=.5,angle=270]{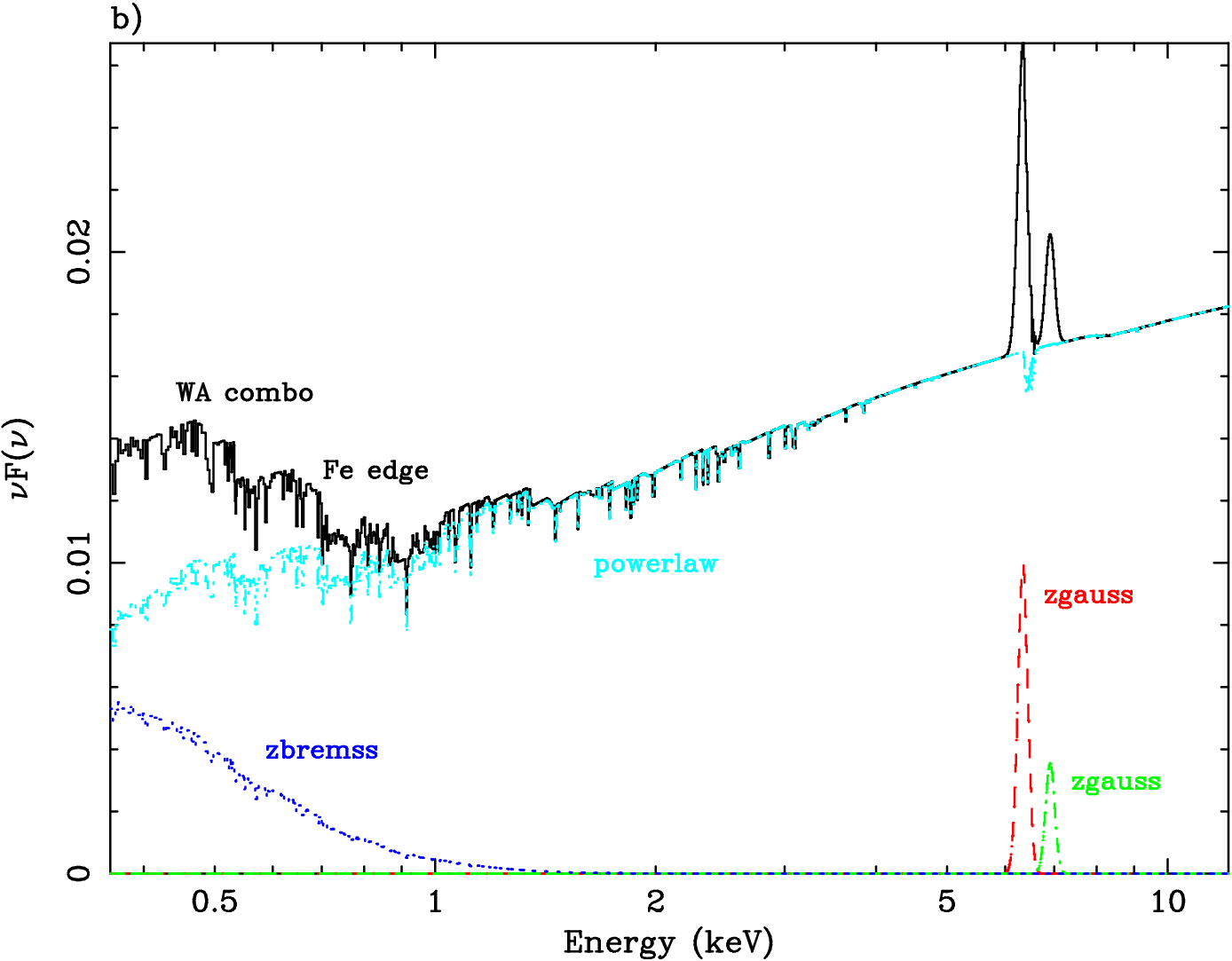}
\caption{(a) The best fit for Model~1, including a {\tt zbremss}
  component to represent the soft emission: $\chi^2/{\rm
  dof}=1813/1745 (1.04)$.  (b) The relative strength
of the model components for Model~1.  The $6.4$ and $6.97 \keV$
Gaussians are shown in red and green, respectively.  The {\tt zbremss}
soft emission is in dark blue, and the photoabsorbed power-law is in
light blue.  Other absorption components are indicated in black.}
\end{figure}

Of course, the description of the soft excess in Model-1 is purely
phenomenalogical in nature.  Ideally, we would like to describe the
soft excess with a model that has a direct physical interpretation.
Initially, we attempt to describe the soft excess as soft X-ray
reflection from a mildly-ionized accretion disk (see Brenneman \&
Reynolds 2006 for a successful application of this model to the
Seyfert galaxy MCG-6-30-15).  Operationally, we replace the thermal
bremsstrahlung component in Model~1 with the ionized disk model {\tt
reflion} created by Ross \& Fabian (2005).  Because the irradiated
matter is also responsible for producing the Fe-K$\alpha$ line in many
sources, this model has the potential advantage of self-consistently
describing the soft excess as well as the 6.4\,keV emission line
feature.  

Interestingly, we find that we {\it cannot} successfully describe both
the soft emission and the Fe-K$\alpha$ line with the disk reflection
model.  The resulting best fit is $\chi^2/{\rm dof}=2652/1745 (1.52)$,
and we found that significant residuals remained on the soft and hard
ends of the spectra.  Even adding the Gaussian component back to
explicitly model the $6.4 \keV$ iron line (and hence removing the
constraint that the disk reflection must describe this feature), the
disk reflection model was unable to adequately fit the form of the
soft excess.  We thus conclude that ionized disk reflection is not an
important process in shaping the soft X-ray spectrum of NGC~4593.
Given the relative narrowness of the Fe-K$\alpha$ line in this source
we believe it is not originating in the inner disk, and hence the
above is not a surprising conclusion.

Alternatively, we can postulate that the soft excess is due to thermal
Comptonization by plasma at a temperature between that of the disk and
the hard X-ray corona (possibly in a transition zone).  Operationally,
we use the {\tt comptt} model (Titarchuk 1994) to model the soft
excess.  Hereafter, we shall refer to this spectral model as Model~2.
This preserves the physical realism of the soft emission arising from
the accretion disk, but allows the iron line to be produced elsewhere,
perhaps farther away in the molecular torus where it would not be as
broad by nature.  Model~2 reaches a best fit of $\chi^2/{\rm
dof}=1759/1744 (1.01)$.  The best fit and model for Model~2 is shown
in Fig.~4.  Here we have frozen the seed photon temperature at $T_0=50
\eV$ and have kept a slab geometry for the corona.  The best fit
Comptonizing plasma temperature is $kT_{\rm c}=42\keV$ with an optical
depth of $\tau=0.12$, though both parameters were not very well
constrained by the fit.  This is not surprising, given that both are
equally involved in shaping the spectrum via the Compton-y parameter:
$y \propto \tau T_{\rm c}$.  We note that for Model~2 the equivalent
widths of the $6.4$ and $6.97 \keV$ Gaussians remain approximately
unchanged from their values in the Model~1 fit.

\begin{figure}
\includegraphics[scale=.5,angle=270]{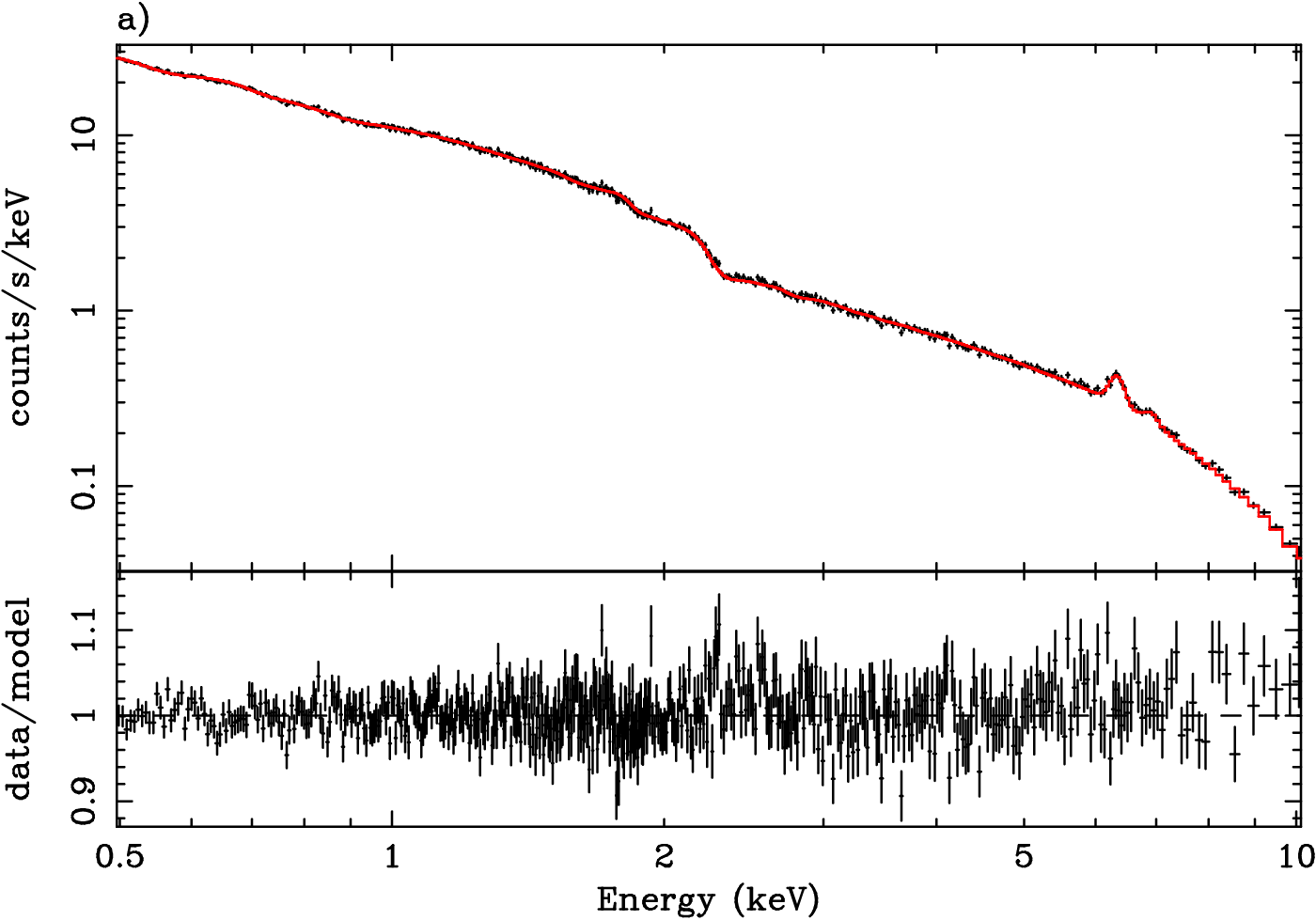}\\
\includegraphics[scale=.5,angle=270]{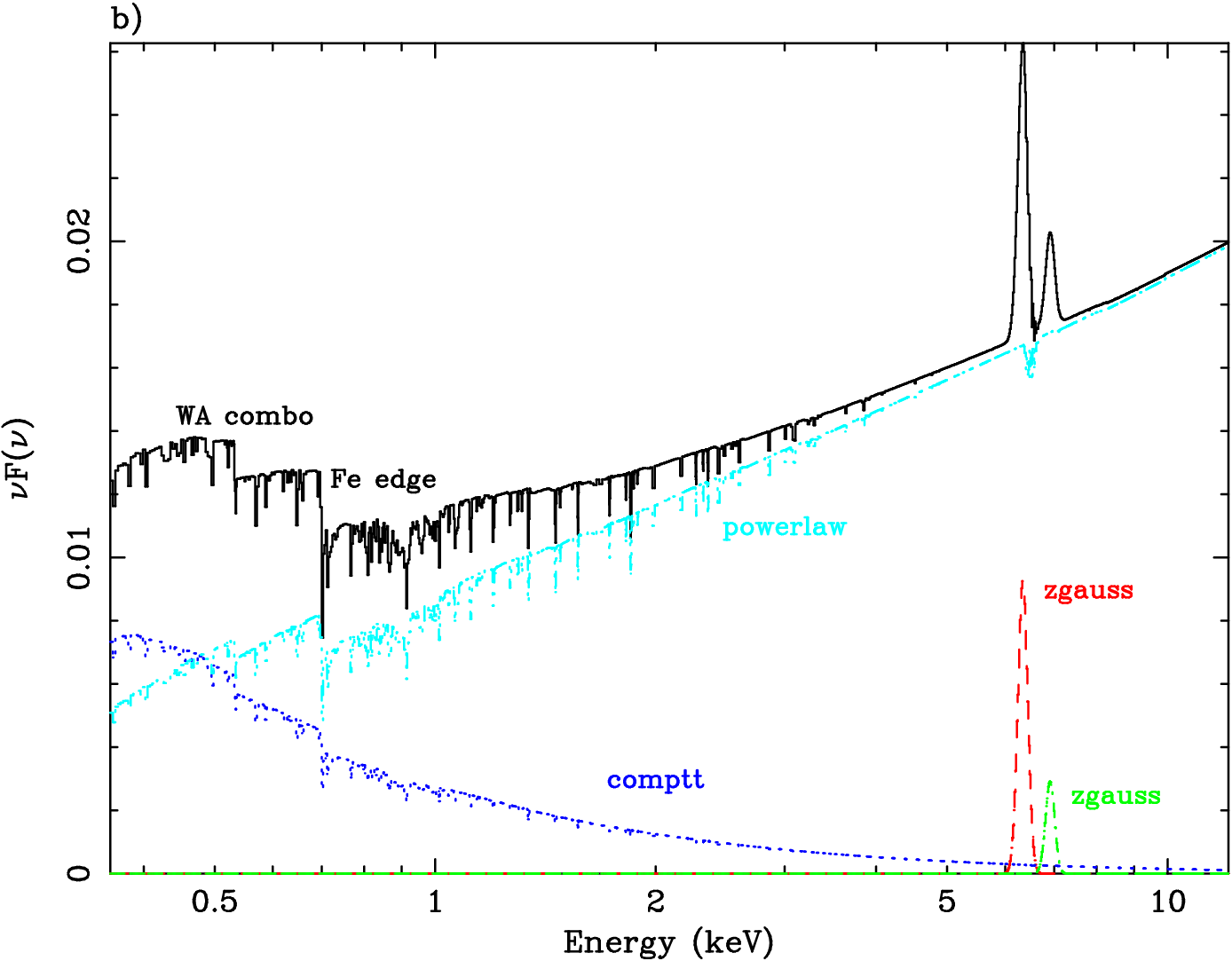}
\caption{(a) The best fit for Model~2, including a {\tt comptt}
  component to represent the soft emission in place of the {\tt
    zbremss} component of Model~1.  $\chi^2/{\rm
  dof}=1808/1744$.  (b) The relative strength
of the model components for Model~2.  The color scheme is the same as
in Fig.~3b.}
\end{figure}

Table~1 reports the parameter values and goodness of fit for both the
final best fitting phenomenalogical model (Model~1, with {\tt zbremss}
and including all other components added in), and the more physical
model (Model~2, including {\tt comptt} and all other components).
Using the Model~2 fit from Table~1, the total $0.5-10 \keV$ flux is
$F_{\rm X}=6.74 \times 10^{-11} \ergpcmsqps$.  Assuming a flat
universe WMAP cosmology, this corresponds to a luminosity of $L_{\rm
X} = 1.21 \times 10^{43} \ergps$.  Considering only the energy range
from $2-10 \keV$, $L_{2-10}=7.40 \times 10^{42} \ergps$.  This is
roughly $21\%$ greater than the $2-10 \keV$ luminosity observed by
McKernan \etal (2003), but only about $86\%$ of the value from the
{\it ASCA} observation reported by Reynolds (1997).

The warm absorber modeled with our two created {\sc xstar} tables
suggests a multi-zone structure in column density and ionization, as
discussed above.  Though our model does not parameterize the covering
fraction of the absorbing gas, we can approximate this fraction by
representing the warm absorber with the O\,{\sc vii} and O\,{\sc viii}
edges at $0.74$ and $0.87 \keV$, respectively, as has been done by
many other authors examining warm absorbers
(e.g., Reynolds 1997; McKernan \etal 2003).  In so doing, we find that
to $90\%$ confidence, the O\,{\sc vii} edge has an optical depth of
$\tau=0.16^{+0.02}_{-0.01}$ whereas the
O\,{\sc viii} edge is much less prominent at $\tau<0.01$.  We therefore
chose to use the O\,{\sc vii} edge as a diagnostic for determining the
covering fraction of the absorbing gas in NGC~4593.  If the O\,{\sc
  vii} edge accounts for the absorption by this gas, the photons
absorbed should be reprocessed and radiated in emission as the O\,{\sc vii}
radiative recombination line at $0.574 \keV$.  Assuming that this line
will be narrow, we fit a Gaussian to the data.  Adding in this
component bettered the overall goodness-of-fit by
$\Delta\chi^2/\Delta{\rm dof}=-53/-1$ as compared to the two-edge
model for the warm absorber, showing that this line is statistically
robust.  The O\,{\sc vii} RRC line had an upper limit equivalent width
of $EW < 6.35 \eV$ and a flux of $2.82 \times 10^{-4} \phpcmsqps$, such
that the line contributed $\sim 7182$ total photons to the model over the total
observation time.  The O\,{\sc vii} edge, by contrast, contributed
$\sim 31991$ total photons.  Taking the ratio of the reprocessed photons to
the absorbed photons, we can estimate a covering fraction for the warm
absorber: $f \approx 0.23$.

\begin{table*}
\begin{center}
\caption{Best fit parameters for the EPIC-pn spectrum}
\begin{tabular} {|l|l|l|l|l|l|l|l|} 
\hline \hline
{\bf Model Component} & {\bf Parameter} & {\bf Model~1 Value} & {\bf
  Model~2 Value} \\
\hline \hline
{\tt phabs} & $N_{\rm H}$ ($\pcmsq$) & $1.97 \times 10^{20}$ & $1.97
\times 10^{20}$ \\
\hline
{\tt WA 1} & $N_{\rm H1}$ ($\pcmsq$) & $1.64^{+0.07}_{-0.09} \times
10^{23}$ & $9.29^{+1.46}_{-9.15}
\times 10^{22}$ \\
              & {\rm log} $\xi_1$ ($\ergcmps$) & $2.54^{+0.03}_{-0.04}$ &
              $2.75^{+0.12}_{-0.32}$ \\
\hline
{\tt WA 2} & $N_{\rm H2}$ ($\pcmsq$) & $2.47^{+0.31}_{-0.73} \times
10^{21}$ & $1.13^{+1.21}_{-0.94}
\times 10^{22}$ \\
           & {\rm log} $\xi_2$ ($\ergpcmps$) & $0.57^{+0.13}_{-0.17}$
           & $1.70^{+0.30}_{-0.59}$ \\
\hline 
{\tt Fe-L$_3$ edge} & {\rm log} $N_{\rm Fe} (\pcmsq)$ &
$16.81^{+0.02}_{-0.05}$ & $16.92^{+0.04}_{-0.08}$ \\
\hline
{\tt po} & $\Gamma$ & $1.87 \pm 0.004$ & $1.75^{+0.02}_{-0.03}$ \\
         & {\rm flux} ($\ergpcmsqps$) & $5.11^{+0.04}_{-0.02} \times 10^{-13}$
         & $4.44^{+0.41}_{-0.54} \times 10^{-13}$ \\
\hline
{\tt zgauss} & $E (\keV)$ & $6.4$ & $6.4$ \\
              & $\sigma (\keV)$ & $0.10 \pm 0.08$ & $0.10 \pm 0.01$ \\
              & {\rm flux} $(\ergpcmsqps)$ & $4.46 \pm 0.32 \times
              10^{-15}$ & $4.15^{+0.33}_{-0.34} \times 10^{-15}$ \\
              & $E (\keV)$ & $6.97$ & $6.97$ \\
              & $\sigma (\keV)$ & $0.10 \pm 0.02$ & $0.10 \pm 0.06$ \\
              & {\rm flux} $(\ergpcmsqps)$ & $9.47^{+2.00}_{-2.07} \times
              10^{-16}$ & $7.77^{+2.07}_{-2.37} \times 10^{-16}$ \\
\hline
{\tt zbremss}  & $kT (\keV)$ & $0.21^{+0.003}_{-0.005}$ & \\
               & {\rm flux} $(\ergpcmsqps)$ & $2.01^{+0.33}_{-0.11} \times 10^{-14}$ & \\
\hline
{\tt comptt}  & $T_0 (\keV)$ & & $0.05$ \\
              & $kT (\keV)$ & & $42.19^{+140.67}_{-40.19}$ \\
              & $\tau_{\rm p}$ & & $0.12^{+0.06}_{-0.11}$ \\
              & {\rm flux} ($\ergpcmsqps$) & & $6.46^{+0.90}_{-0.92} \times 10^{-14}$ \\
\hline
{\bf $\chi^2/{\rm dof}$} & & $1813/1745 (1.04)$ & $1759/1744 (1.01)$ \\
\hline \hline
\end{tabular}
\end{center}
\label{tab:tab1.tex}
\small{The energy range from $0.5-10.0 \keV$ is considered for the
  EPIC-pn.  Best fit Model~1 contains a {\tt zbremss} component to parameterize
  the soft excess below $\sim 2 \keV$, while best fit Model~2 represents this
  component with a {\tt comptt} model.  All quoted error bars are at
  the $90\%$ confidence level.  All redshifts used were frozen at the 
  cosmological value for NGC~4593: $z=0.009$.  }
\end{table*}

\section{Variable Nature of the Source}
\label{sec:variability}

NGC~4593 displays significant variability (over a factor of two in
$0.5-10 \keV$ flux) during the course of this observation.  In this
Section, we examine the detailed variability properties of this
source.

\subsection{Spectral Variability}
\label{sec:spec_var}

We have analyzed the spectral variability of the source as seen in the
EPIC-pn data.  As an initial assessment of spectral variability,
Fig.~5 plots the data-to-model ratios of consecutive $10 \ks$ segments
of the EPIC-pn data using the {\tt comptt} best fit spectrum model
discussed in \S3.  Experimentation suggested that intervals smaller
than $10 \ks$ contain insufficient counts to maintain the integrity of
the spectrum.  As well as the obvious variations in the normalization
of the spectrum, there are clear changes in slope between segments
(with the source becoming softer as it brightens).  There is also a
feature of variable equivalent width seen at $\sim 6.4 \keV$,
corresponding to the cold Fe-K$\alpha$ line.  Interestingly, though,
we see no variable discrete features in the soft X-ray spectrum
suggesting that there is no significant warm absorber variability
during our observations.

To examine variability of the iron lines in more detail, Fig.~6 plots
the renormalized interval data over the time-averaged data from $2-10
\keV$.  Direct spectral fitting of the the cold and ionized Fe-K lines
in each interval with Gaussian profiles illuminates the nature of
their variability (or lack thereof).  As can be seen in Table~2 and
Figs.~7, the cold line exhibits an approximately constant flux and a
variable equivalent width.  The constant equivalent width hypothesis
is rejected with greater than 99\% significance, whereas a constant
flux hypothesis is consistent with the data at the 20\% level.  Thus
the evidence suggests that the cold iron line flux does not appear to
respond to changes in the X-ray continuum.  Constraints on the ionized
line are not strong enough to rule out either a constant flux and
equivalent width, as is shown in Fig.~8 and Table~2.

\begin{figure}
\centerline{
\includegraphics[scale=.3,angle=270]{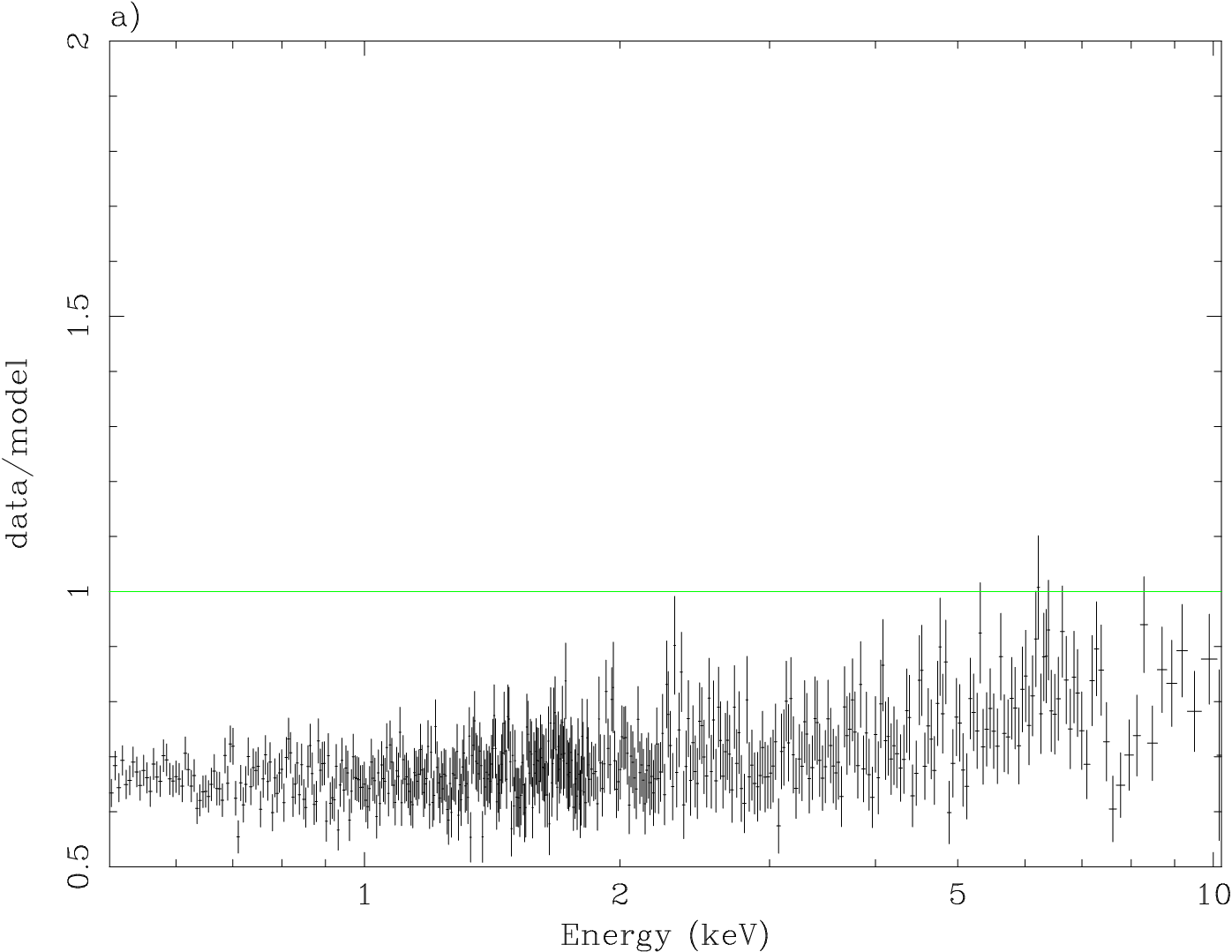}
\includegraphics[scale=.3,angle=270]{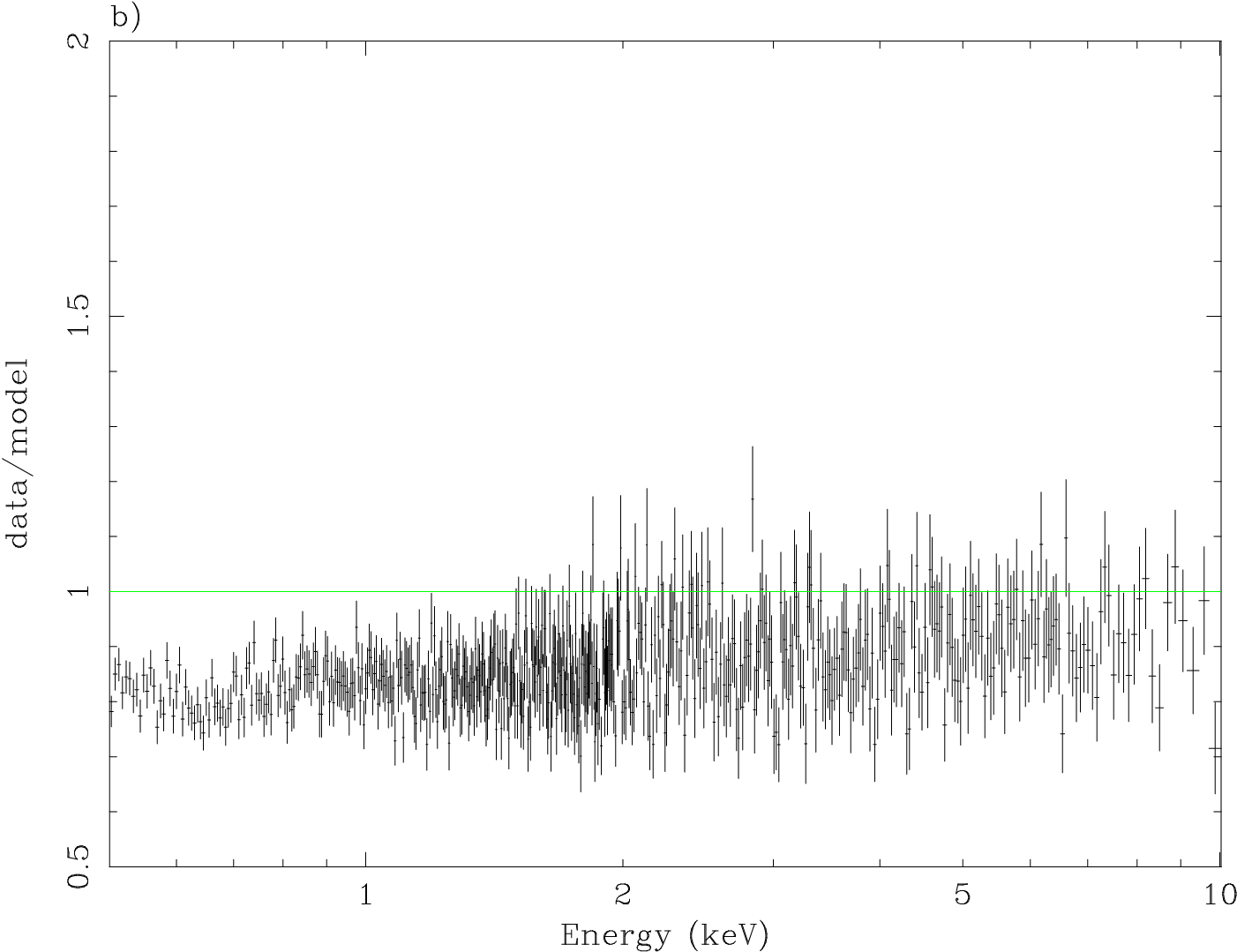}
}
\centerline{
\includegraphics[scale=.3,angle=270]{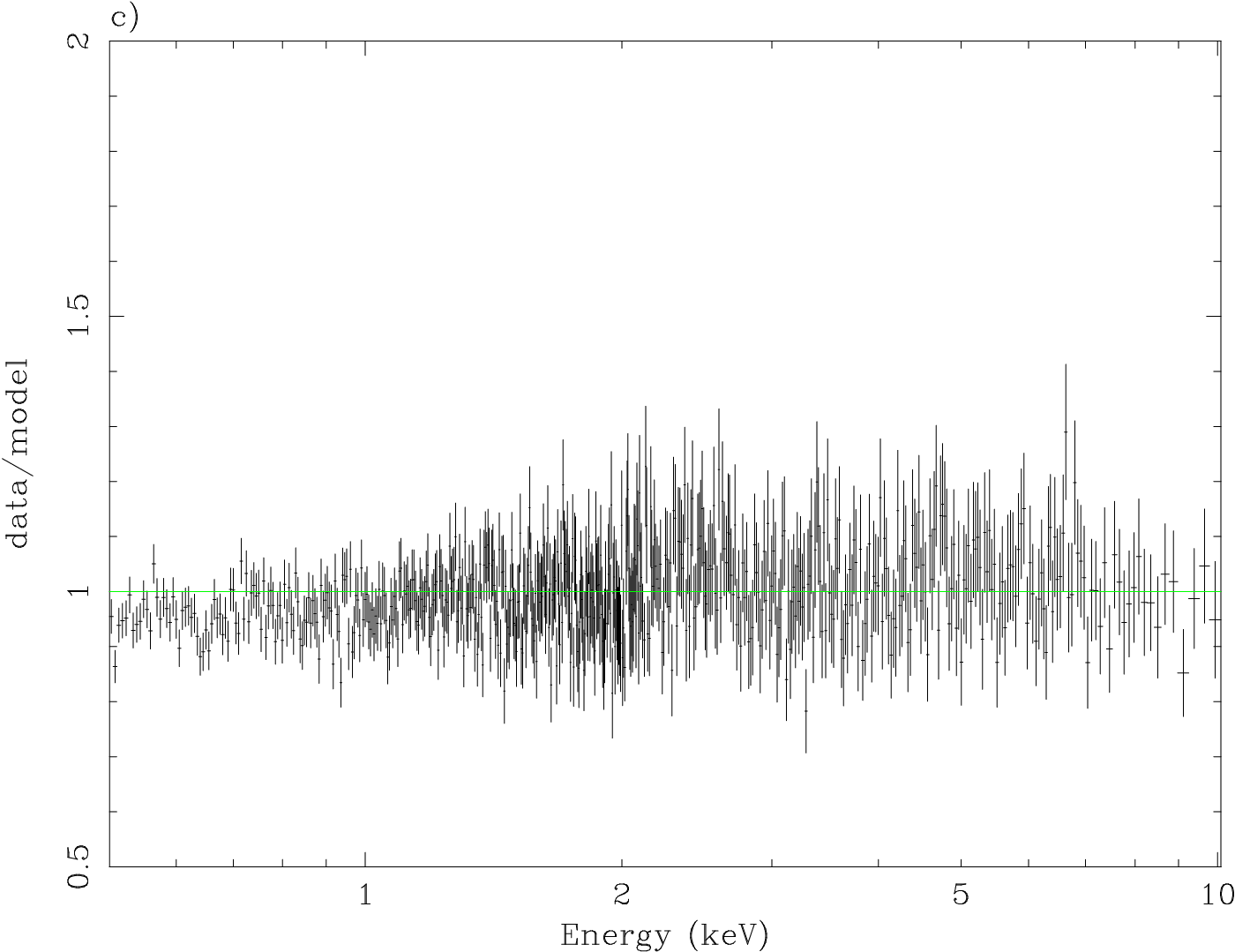}
\includegraphics[scale=.3,angle=270]{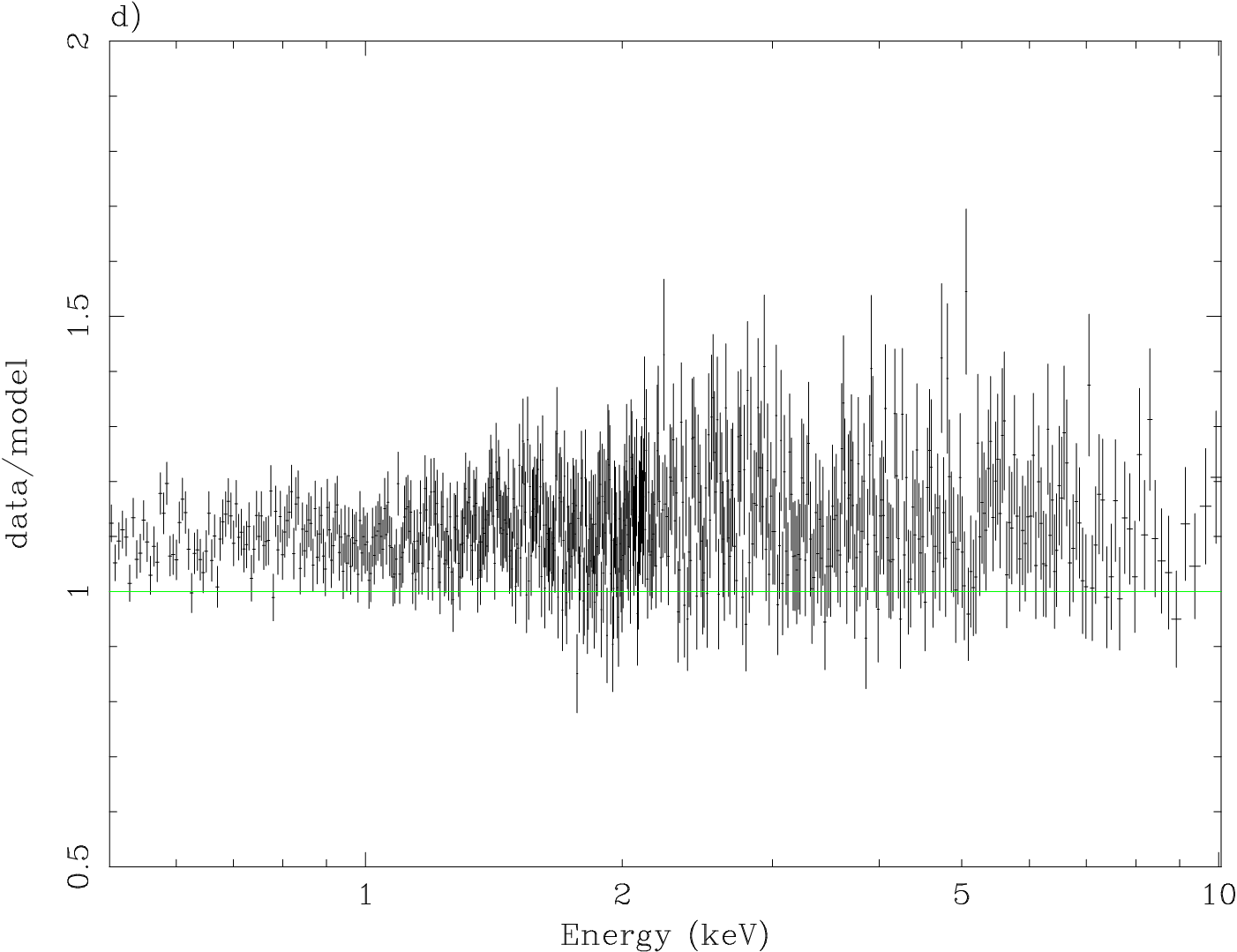}
}
\centerline{
\includegraphics[scale=.3,angle=270]{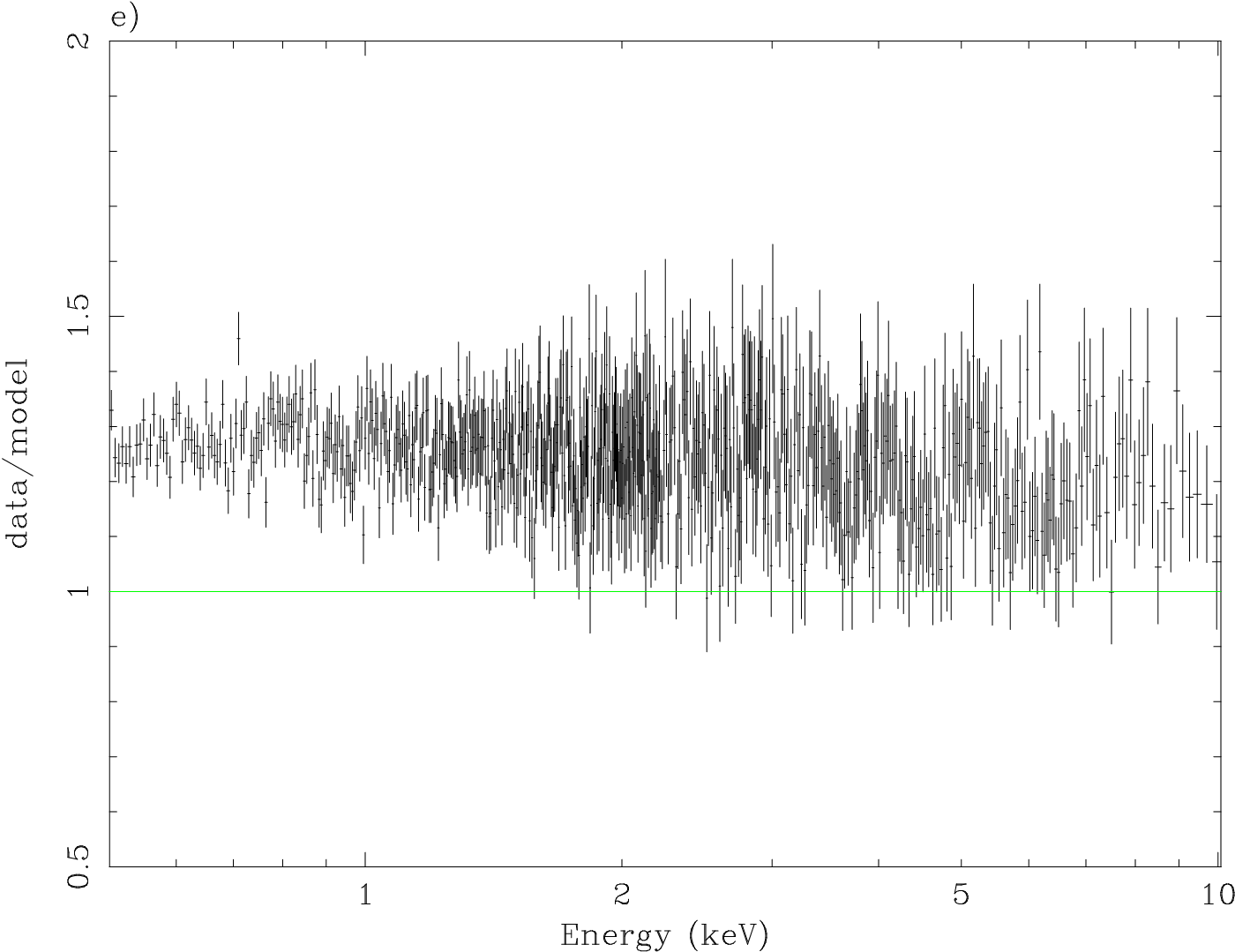}
\includegraphics[scale=.3,angle=270]{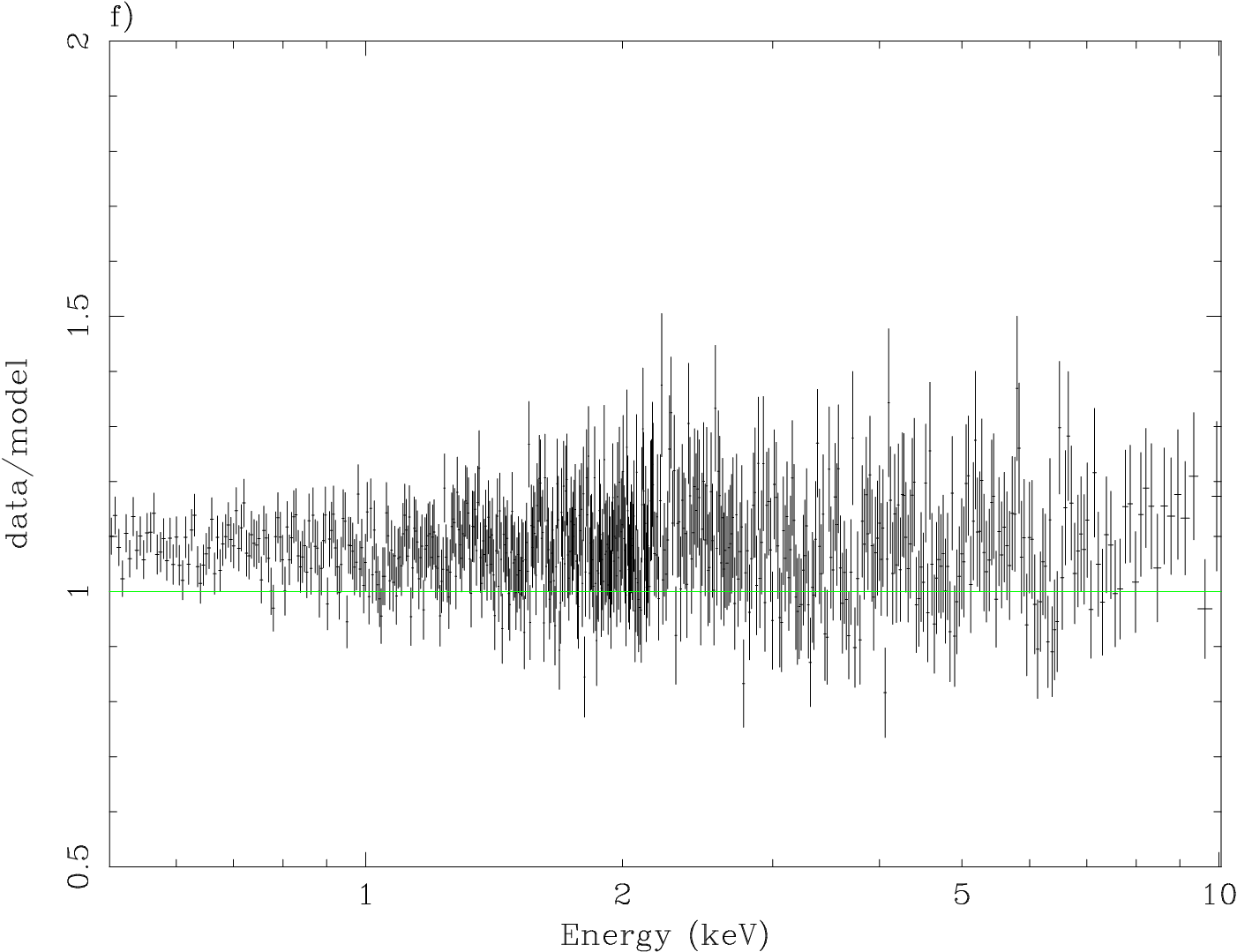}
}
\centerline{
\includegraphics[scale=.3,angle=270]{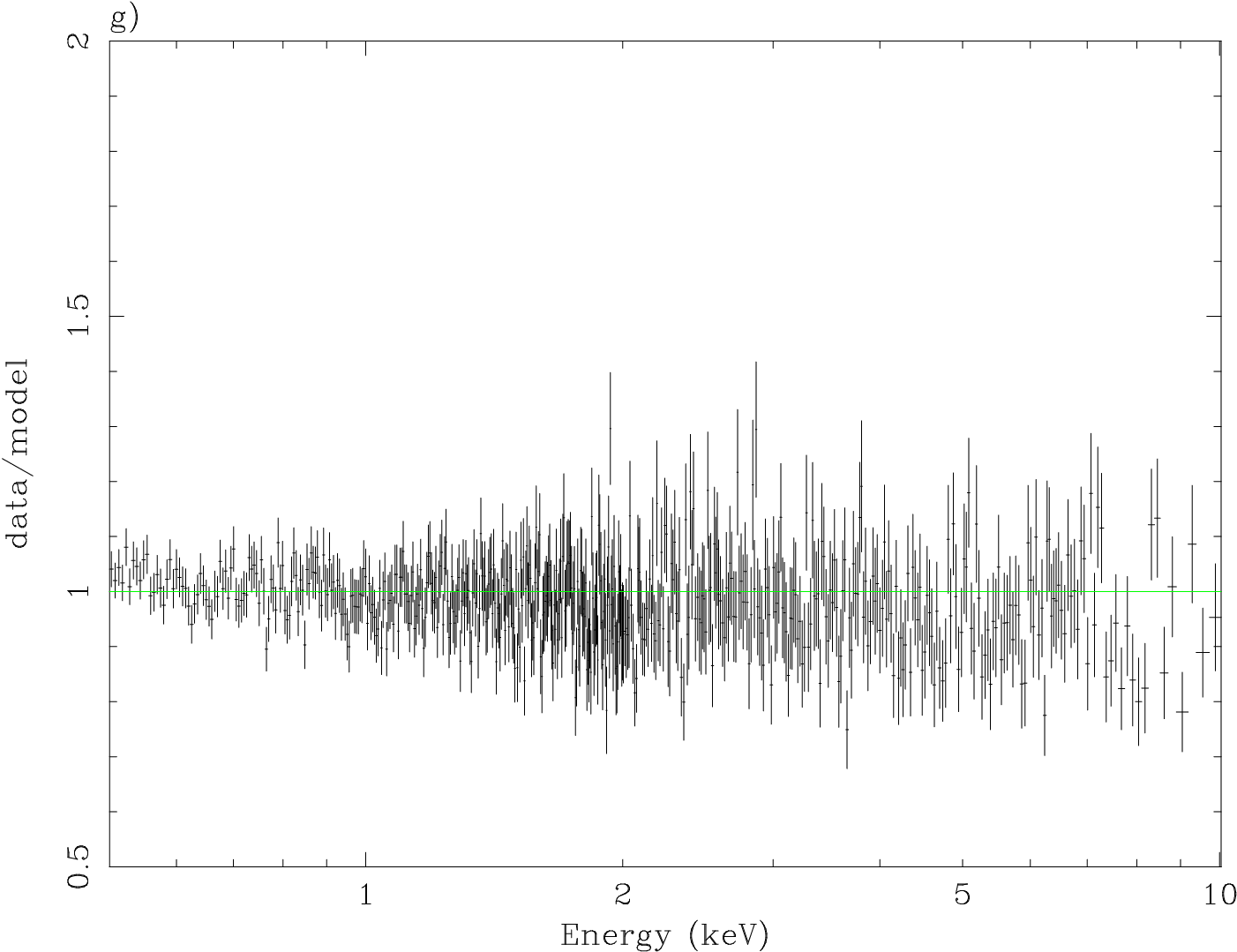}
\includegraphics[scale=.3,angle=270]{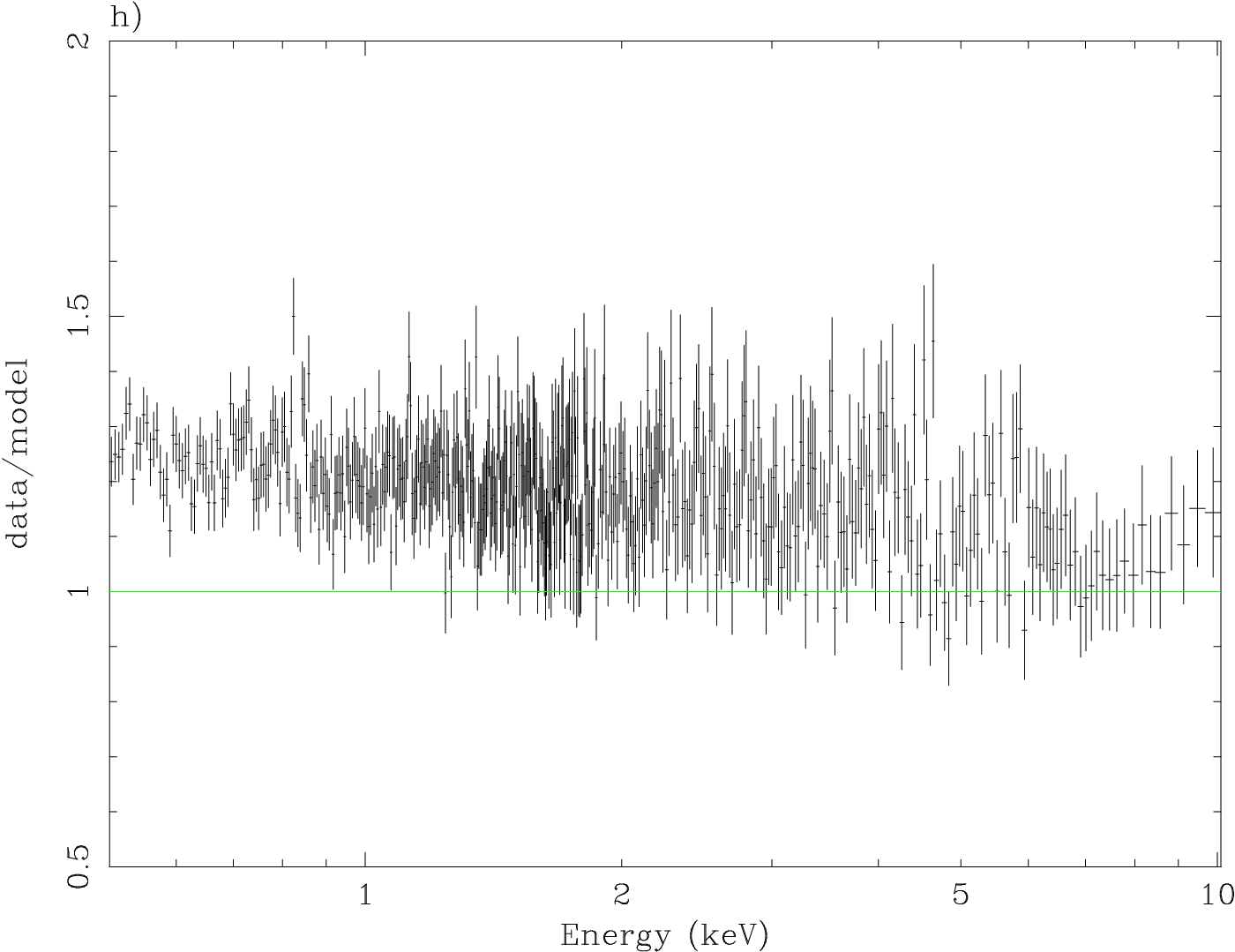}
}
\caption{Variation of the EPIC-pn time-averaged spectrum from the 
{\tt comptt} best fit model discussed in \S3. 
Time intervals are $10 \ks$ in length except for the last, which is
$\sim 6100 \ks$.  Segments a) - h) are shown in chronological order 
of the observation.}
\end{figure}   

\begin{figure}
\centerline{
\includegraphics[scale=.3,angle=270]{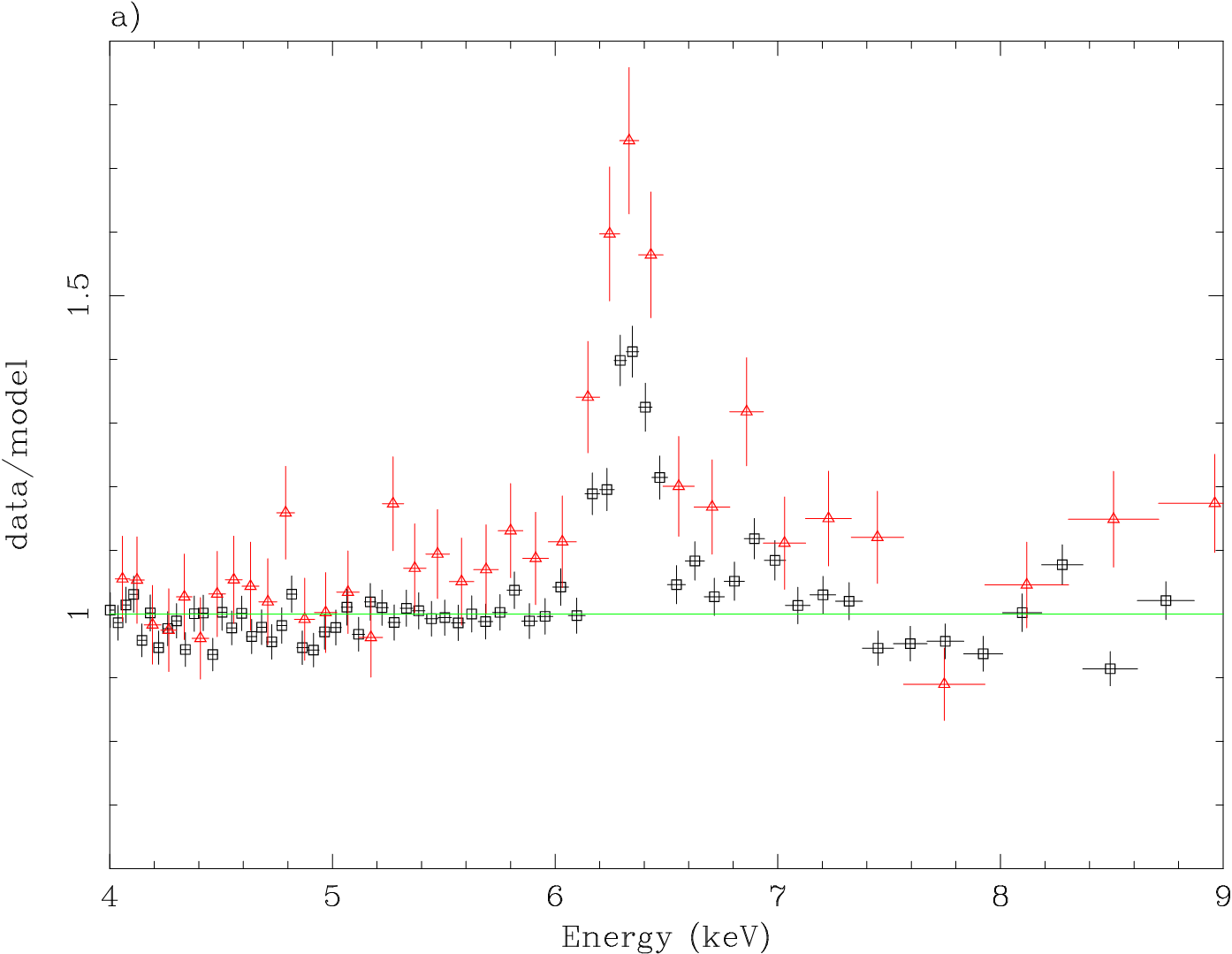}
\includegraphics[scale=.3,angle=270]{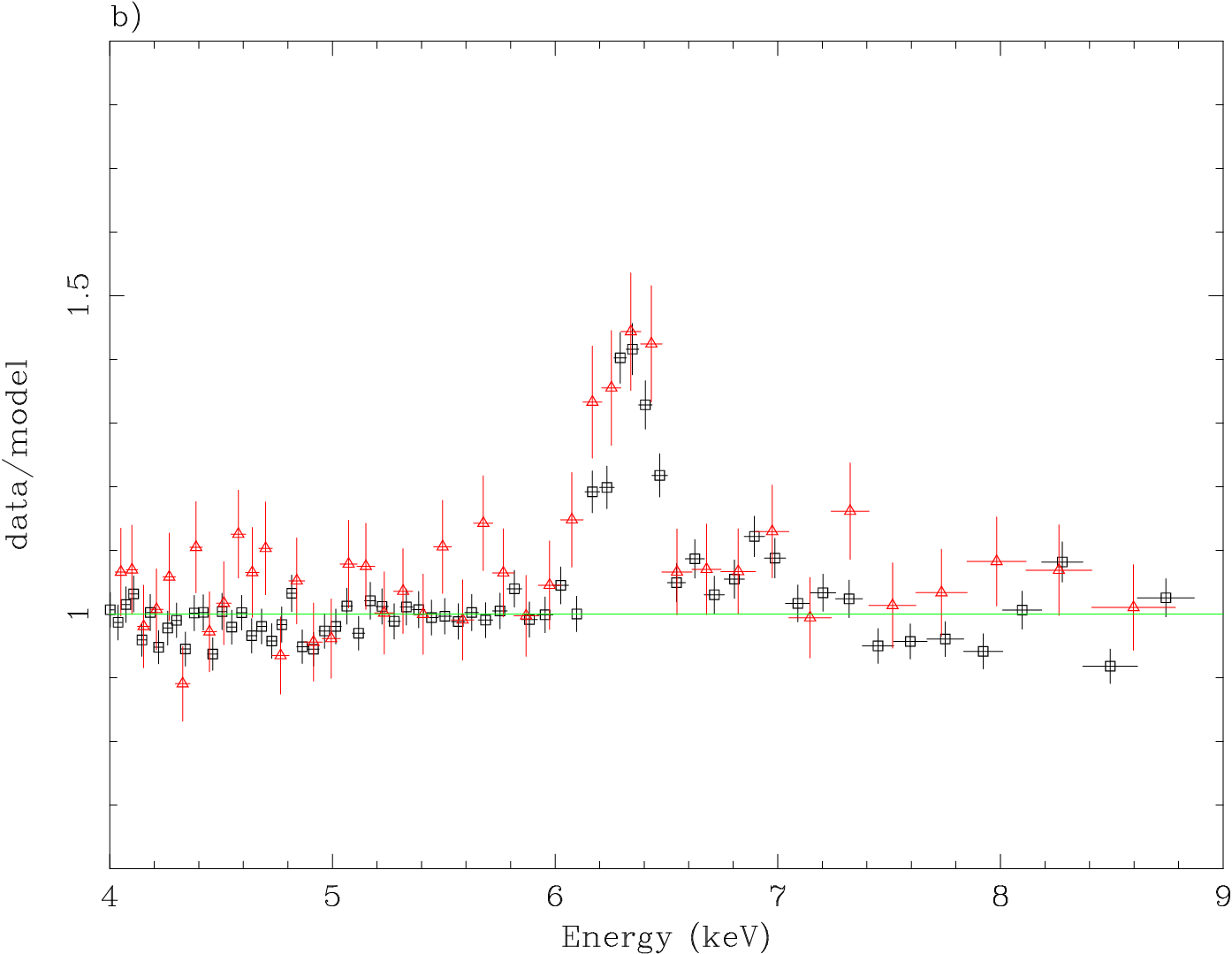}
}
\centerline{
\includegraphics[scale=.3,angle=270]{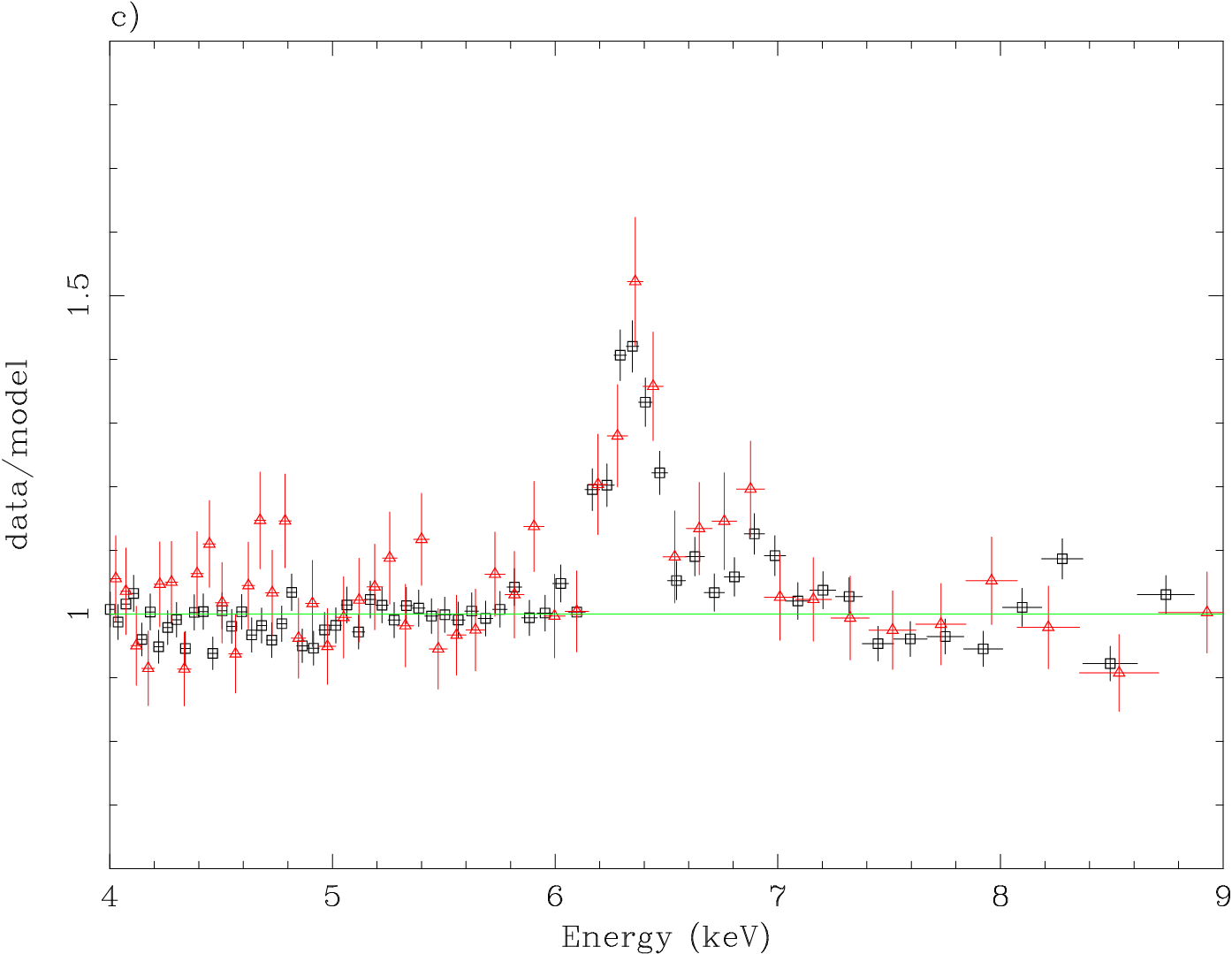}
\includegraphics[scale=.3,angle=270]{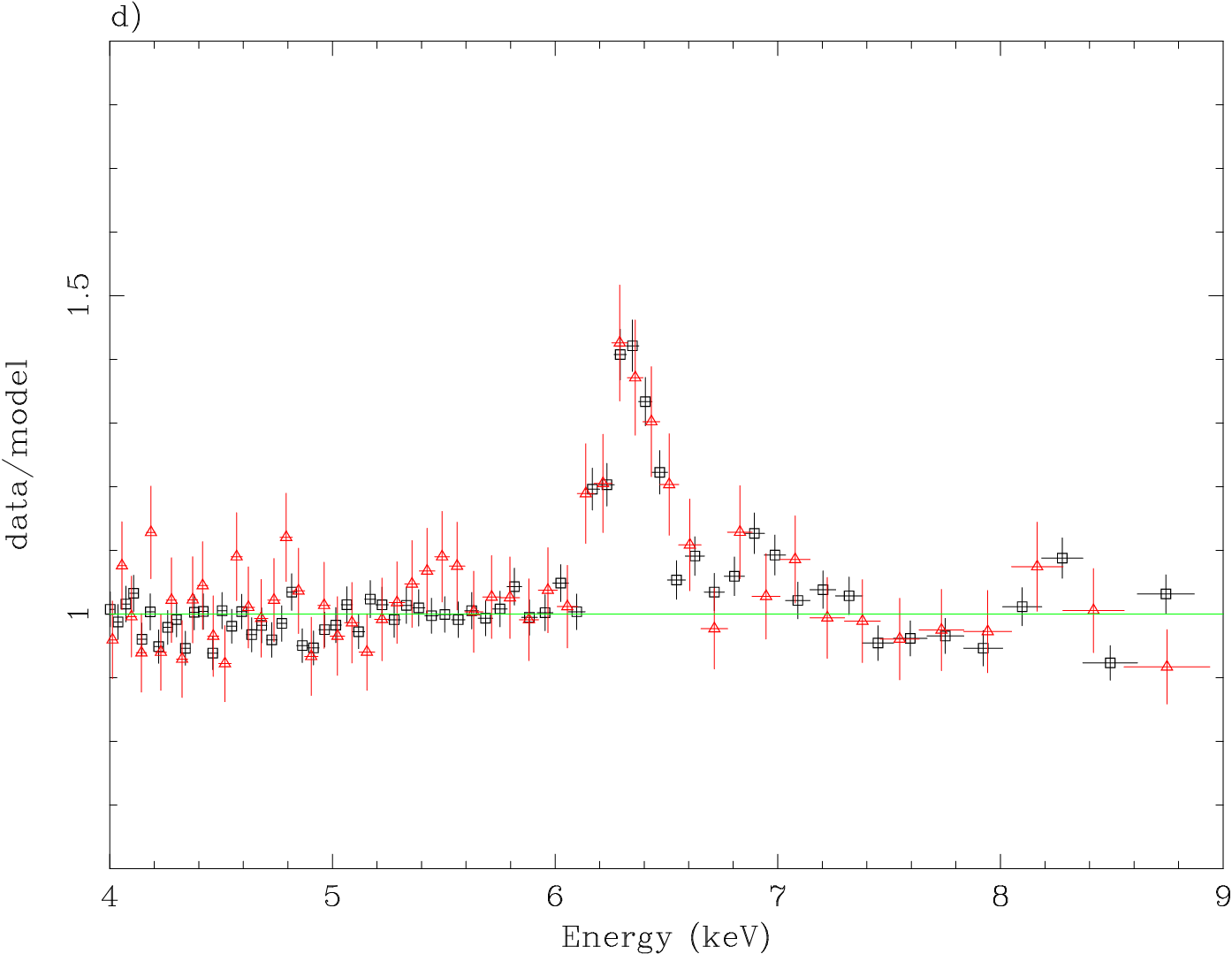}
}
\centerline{
\includegraphics[scale=.3,angle=270]{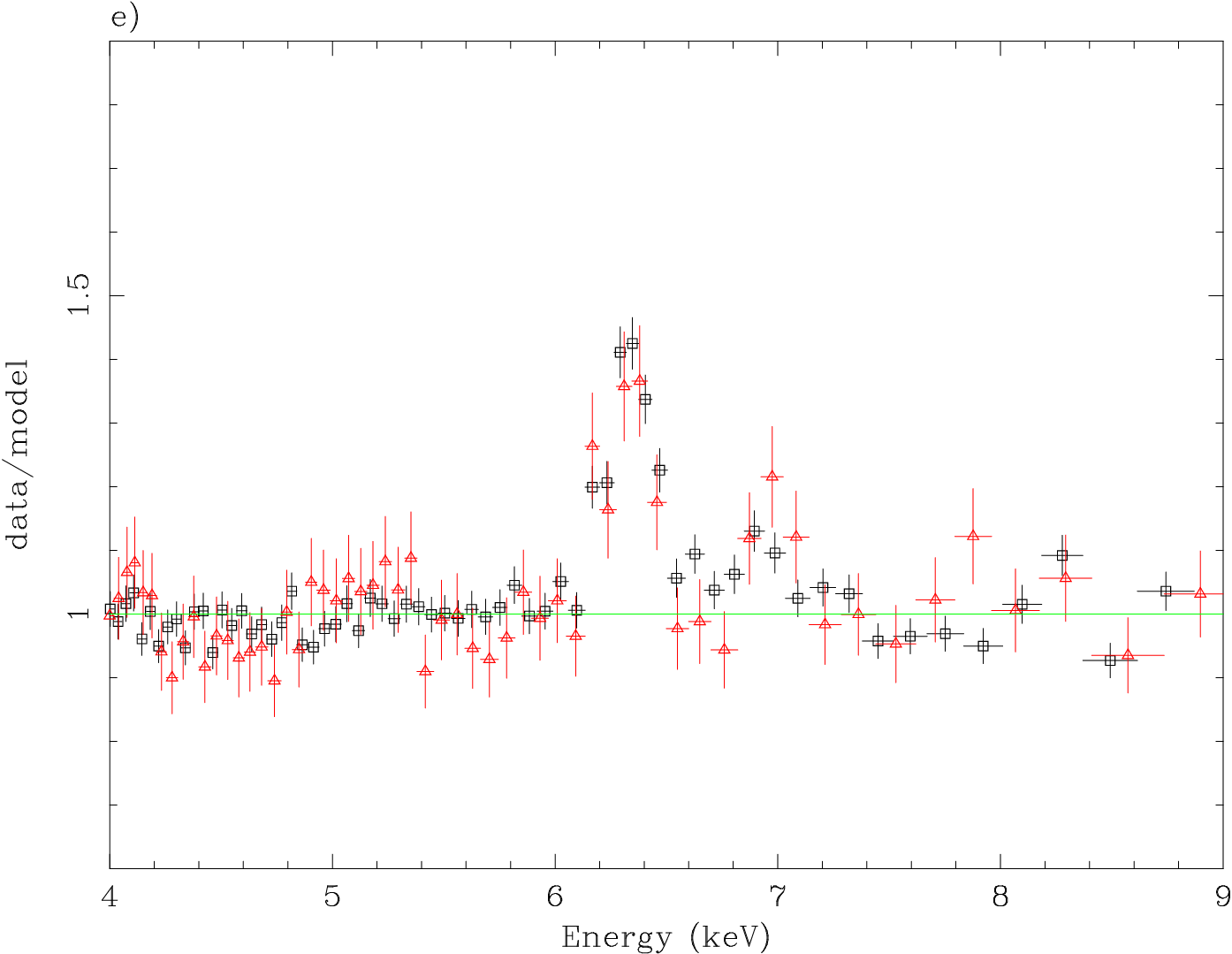}
\includegraphics[scale=.3,angle=270]{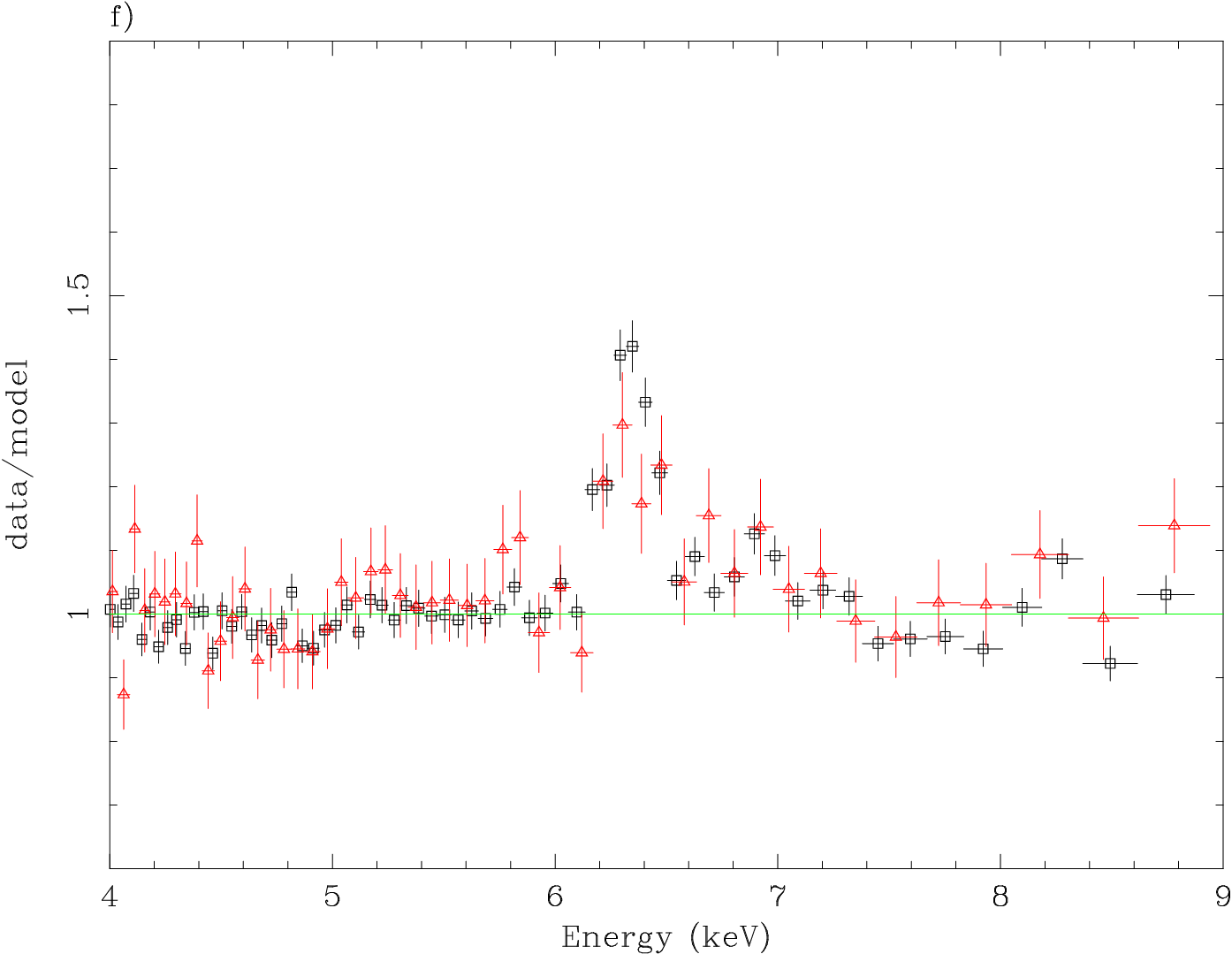}
}
\centerline{
\includegraphics[scale=.3,angle=270]{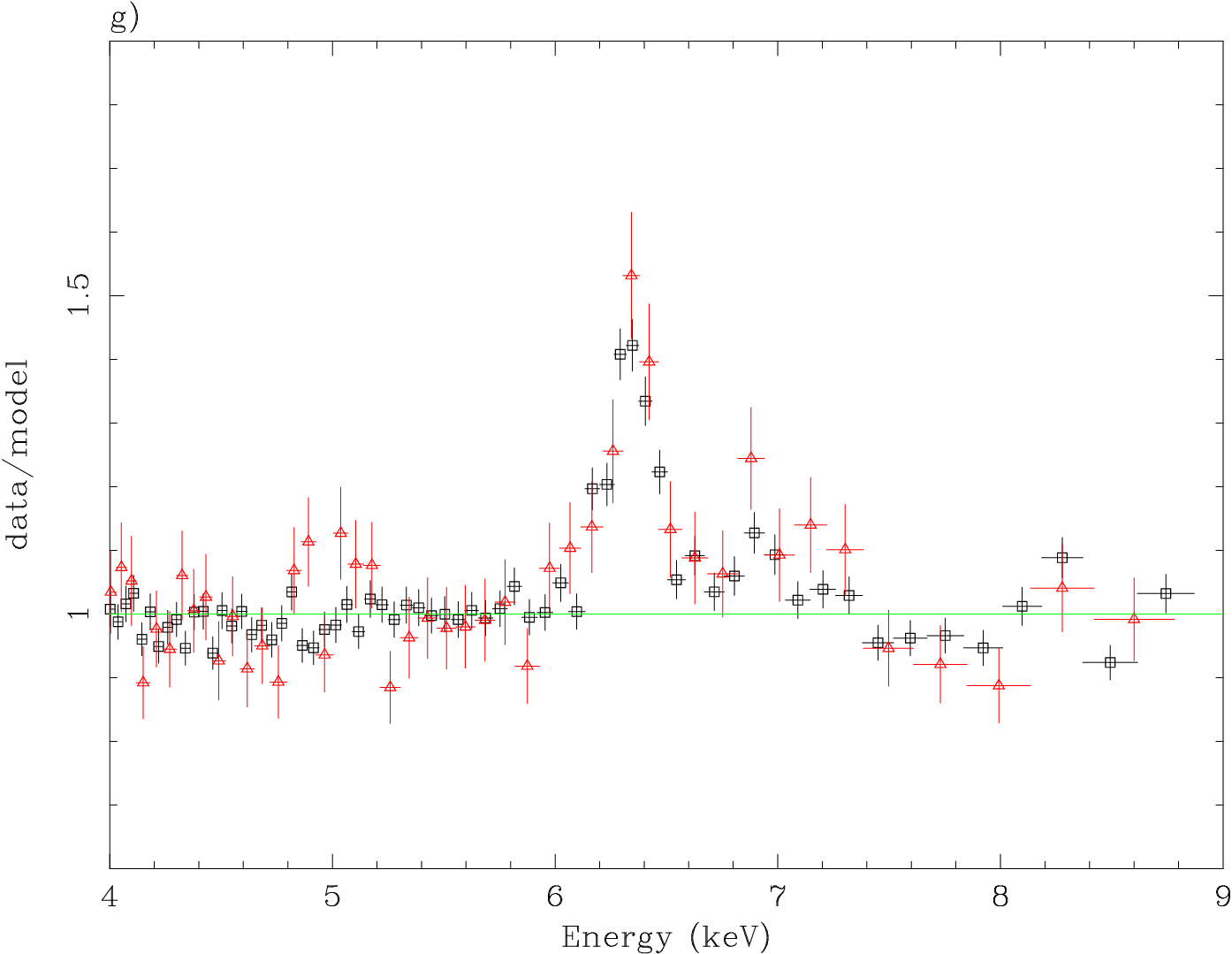}
\includegraphics[scale=.3,angle=270]{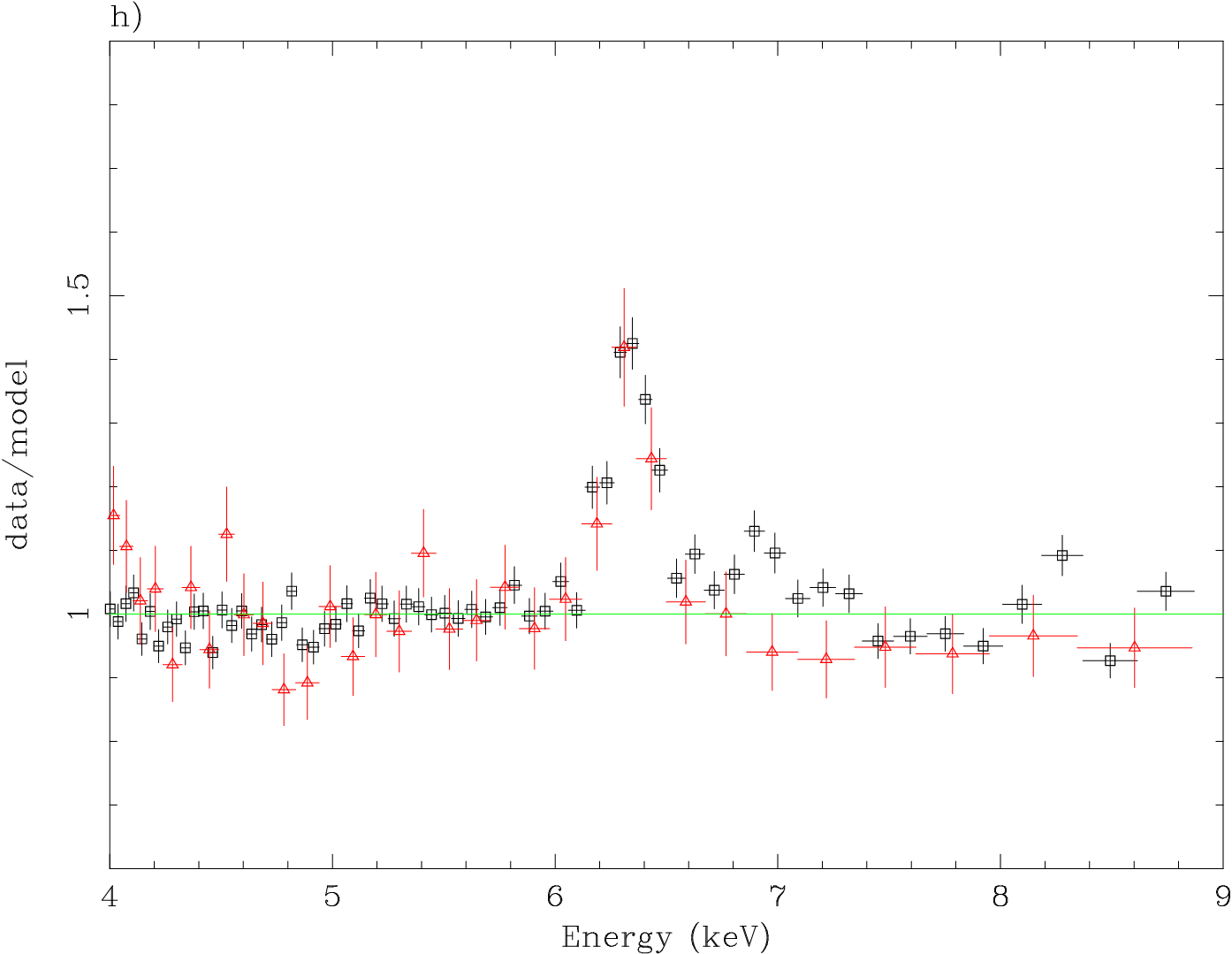}
}
\caption{The cold and ionized Fe-K$\alpha$ lines from the
  time-separated EPIC-pn spectrum.  Time intervals are $10 \ks$ each
  in length, as above in Fig.~5.  Lines are above a {\tt
  phabs po} continuum.  The time-averaged data appear in black squares,
  interval data in red triangles.}
\end{figure}

\begin{table*}
\begin{center}
\caption{The cold and ionized iron 
lines in the EPIC-pn spectrum of NGC~4593}  

\begin{tabular}{|l|l|l|l|l|} \hline \hline
{\bf Line} & {\bf FWHM ($\kmps$)} & {\bf Flux ($\ergpcmsqps$)} & {\bf EW ($\eV$)} \\
\hline
Cold Fe-K$\alpha$ & $13245 \pm 2208$ & $7.41 \pm 0.86 \times 10^{-13}$ & $232 \pm 27$ \\
($6.40 \keV$)     & $8831 \pm 2208$ & $4.98 \pm 0.89 \times 10^{-13}$ &$126 \pm 22$ \\
                  & $9934 \pm 2208$ & $5.76 \pm 0.85 \times 10^{-13}$ &$134 \pm 19.8$ \\
                  & $13246 \pm 3311$ & $7.26 \pm 1.04 \times 10^{-13}$& $153 \pm 21.8$ \\
                  & $8831 \pm 2208$ & $5.27 \pm 0.86 \times 10^{-13}$& $105 \pm 17.2$ \\
                  & $11038 \pm 4415$ & $4.08 \pm 0.93 \times 10^{-13}$& $88.5 \pm 20.2$ \\
                  & $8831 \pm 2208$ & $5.61 \pm 0.88 \times 10^{-13}$ & $137 \pm 21.5$ \\
                  & $11038 \pm 3311$ & $5.82 \pm 1.17 \times 10^{-13}$ & $124 \pm 24.9$ \\
\hline
Ionized Fe-K$\alpha$ & $7095 \pm 6081$ & $1.97 \pm 1.41 \times 10^{-13}$ & $64.3 \pm 46.0$ \\
($6.97 \keV$)        & $50678 \pm 83111$ & $2.89 \pm 5.05 \times 10^{-13}$ & $80.3 \pm 140$ \\
                     & $6081 \pm 6081$ & $2.09 \pm 1.54 \times 10^{-13}$ & $51.9 \pm 38.3$ \\
                     & $14190 \pm 11149$ & $2.04 \pm 2.22 \times 10^{-13}$ & $45.6 \pm 49.6$ \\
                     & $9122 \pm 6081$ & $2.52 \pm 1.85 \times 10^{-13}$ & $53.7 \pm 39.4$ \\
                     & $18244 \pm 9122$ & $3.06 \pm 3.48 \times 10^{-13}$ & $71.3 \pm 57.8$ \\
                     & $24325 \pm 9122$ & $4.90 \pm 3.00 \times 10^{-13}$ & $132 \pm 80.9$ \\
                     & ------- & ------- & ------- \\
\hline \hline
\end{tabular}
\end{center}
\label{tab:tab2.tex}
Each row represents a time 
interval of $10 \ks$ in the observation.  Fits were done in {\sc xspec} 
using a {\tt phabs po} model for the continuum and two Gaussian lines 
with parameters fit to the data in each time interval.  Error bars are 
at $1\sigma$. The blank lines in the ionized
Fe-K$\alpha$ table represent time intervals for which a robust fit to
the data could not be achieved.
\end{table*}

\begin{figure}
\includegraphics[scale=.5,angle=270]{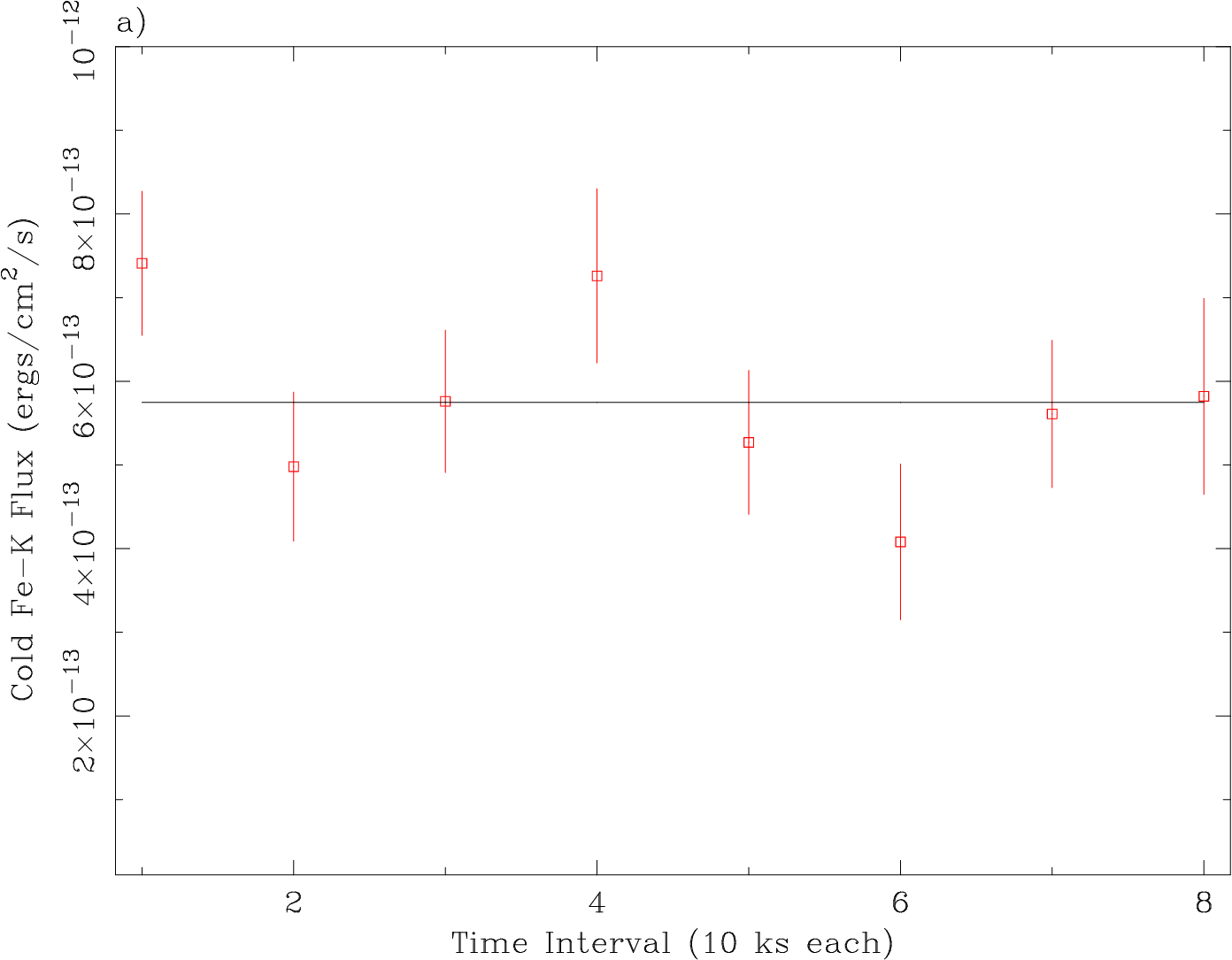}\\
\includegraphics[scale=.5,angle=270]{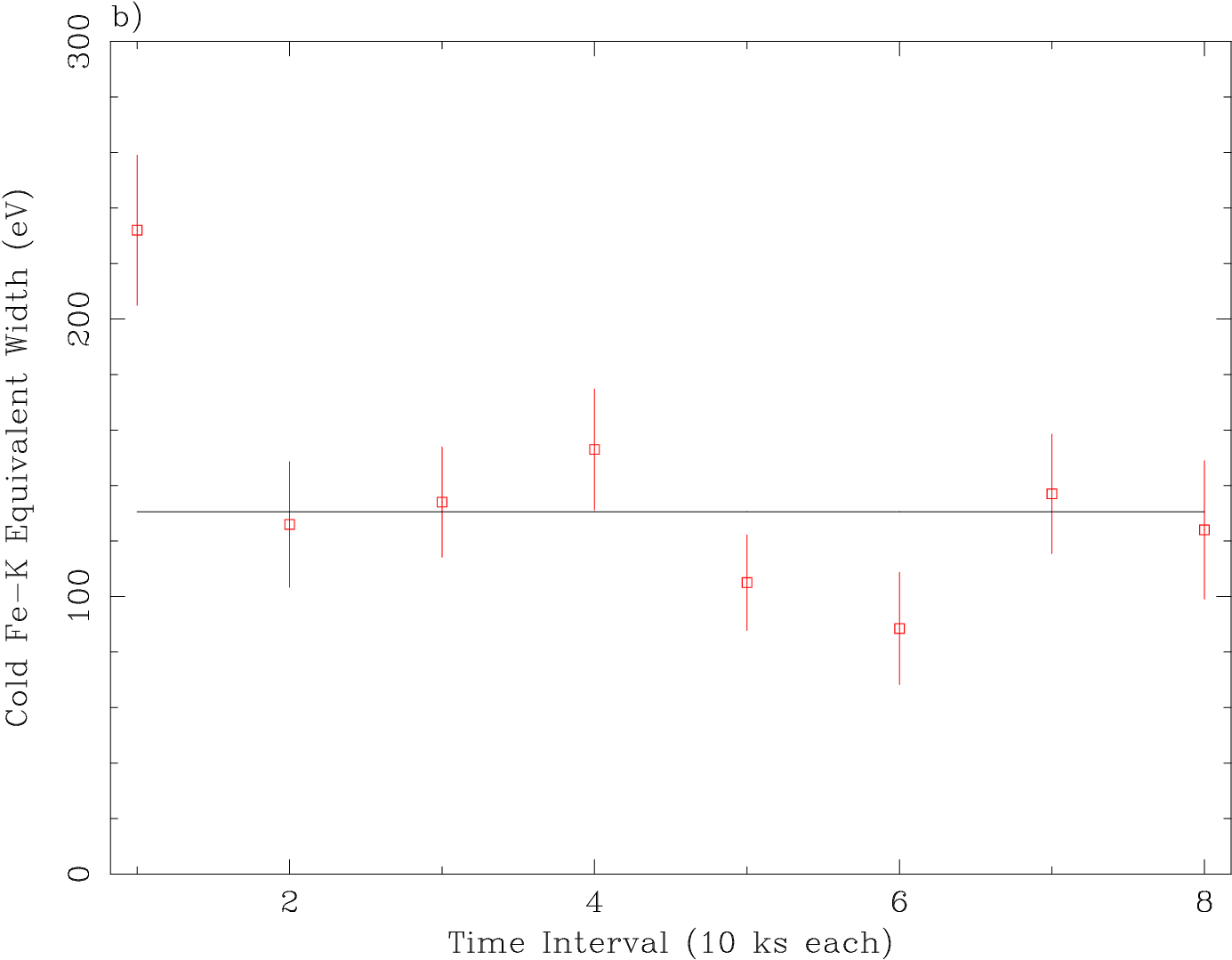}
\caption{Variation of the flux (a) and equivalent width (b) of the
  cold Fe-K$\alpha$ line ($6.4 \keV$) in $10 \ks$ intervals over the
  course of the $76 \ks$ observation.  A fit to the data with a
  constant flux model yields $\chi^2/{\rm dof}=10/7$ ($1.43$) implying
  a 20\% probability that the data are consistent with the constant
  flux model.  The constant equivalent width fit yields $\chi^2/{\rm
  dof}=22/7$ ($3.14$) allowing the constant equivalent width
  hypothesis to be rejected at more than the 99\% significance level.}
\end{figure}   

\begin{figure}
\includegraphics[scale=.5,angle=270]{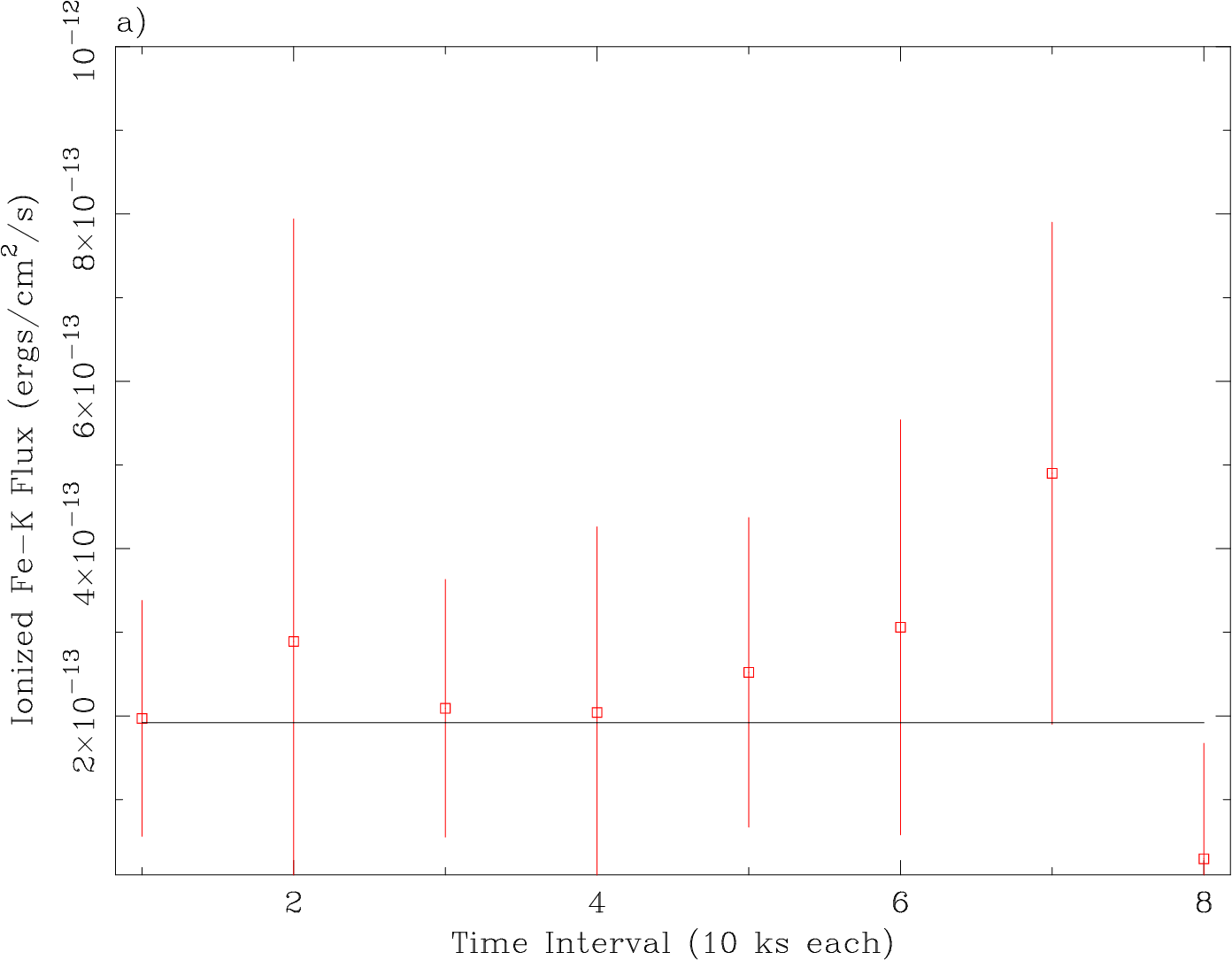}\\
\includegraphics[scale=.5,angle=270]{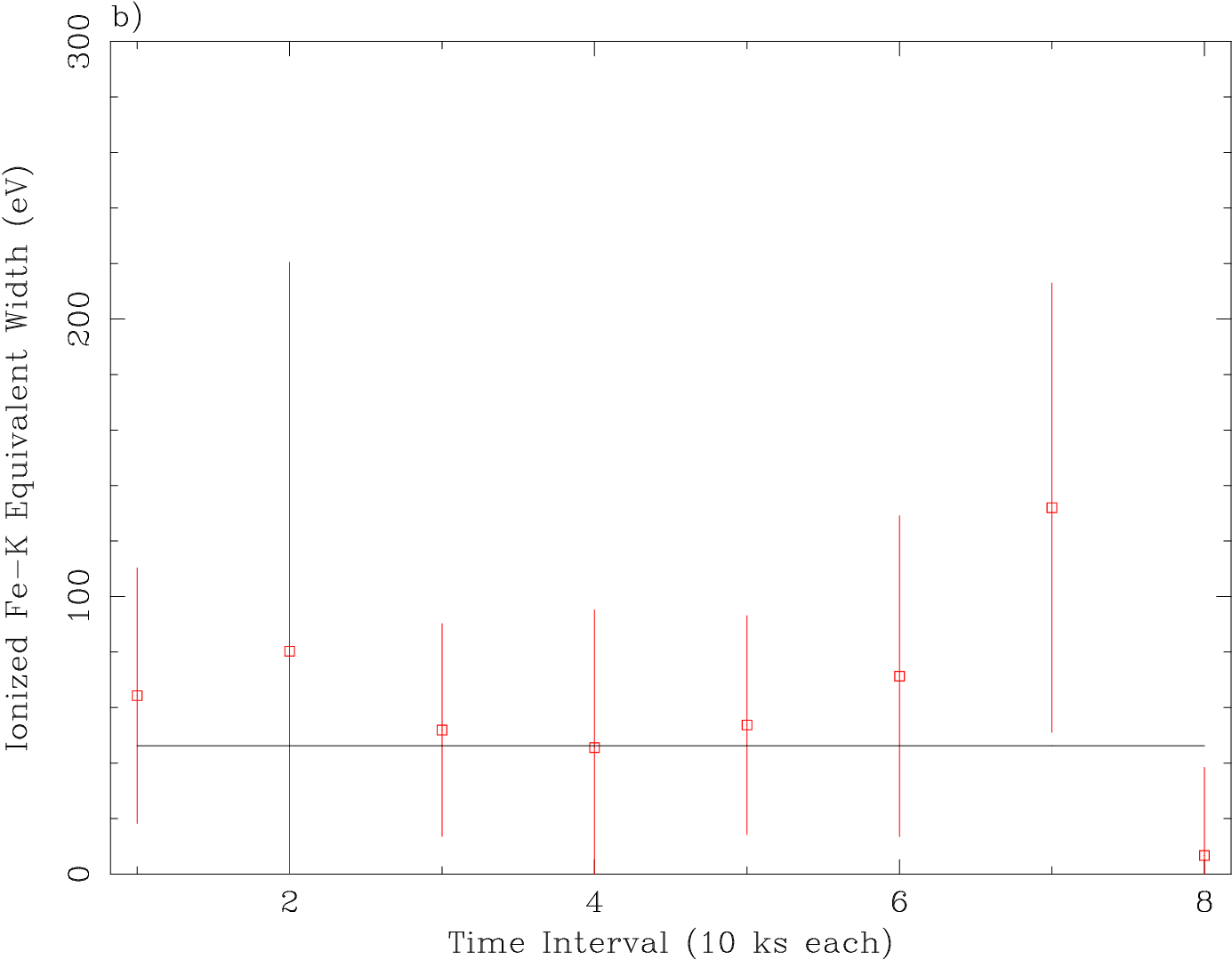}
\caption{Variation of the flux (a) and equivalent width (b) of the ionized Fe-K$\alpha$
  line ($6.97 \keV$) over the course of the observation.  Here the
  uncertainty in the data is so great that a constant flux 
  cannot be ruled out: $\chi^2/{\rm dof}=3/7$ ($0.43$).  A constant equivalent
  width is similarly likely: $\chi^2/{\rm dof}=3/7$ ($0.43$).}
\end{figure}   

The simplest way to interpret the lack of continuum response of the
cold iron line is to suppose that it exists on spatial scales with
light travel times greater than the duration of our observation.  For
an observation time of $76 \ks$, as is the case here, the light would
travel approximately $150$ AU, or roughly $1.9\times 10^3\,r_g$ if NGC
4593 harbors a $8.1\times 10^6 \Msun$ black hole at its core (see
discussion at the end of \S4.2).  The lack of response for the
iron line flux would suggest that this size is a lower limit for the
size of the line emitting source.  This would place the line emission
in the outer regions of the accretion disk or the putative molecular
torus of the Seyfert unification scheme.

\subsection{Continuum Power Spectrum and Spectral Lags}
\label{sec:temp_var}

The light curve for the full-band (0.5-10.0 \keV) data set is shown in
Fig.~9a.  The light curve is characterized by an
initial rapid drop up to $\sim 5\ks$, followed by a relatively
steady increase peaking at $\sim 45 \ks$, a drop until $\sim 65 \ks$,
and then a final increase that shows possible signs of tapering off
around the end of the observation at $76 \ks$.  There are indications
of smaller scale variability on time scales of $< 1000 \s$, with
slight flares and drops occurring throughout the data set.

Fig.~9b shows the Leahy normalized power spectrum of the $0.5-10 \keV$
EPIC-pn light curve.  The power spectrum demonstrates that the
variability of the source becomes dominated by Poisson statistics on
timescales of about $200 \s$ once it drops to a power of $\sim 2$, as
expected.  Before this point, the slope of the power spectrum for low
frequencies is $-2.63^{+0.66}_{-0.25}$.  This value for the PSD slope
is consistent with results from several other Sy-1 AGN such as NGC
3783 (Markowitz 2005), MCG--6-30-15 (Papadakis \etal 2005; Vaughan
\etal 2003), NGC 4051 (at high frequencies; McHardy \etal 2004), and
Mrk 766 (also at high frequencies; Vaughan \& Fabian 2003). 

Given the range of X-ray energies available in this dataset, it is
also useful to study the continuum variability in different energy
bands.  Hardness ratios and time lags between different energy bands
in the continuum, in particular, can help constrain the emission
mechanisms of the source as well as the physical scale of the emitting
region(s).  We divide the X-ray spectrum into three energy bands,
each with an approximately equal number of counts ($\sim 5 \times
10^5$).  The soft band ranges from $0.5-0.84 \keV$, the mid band
extends from $0.84-1.5 \keV$, and the hard band covers $1.5-10.0
\keV$.  The light curves for all three bands are represented in
Fig.~10.  In order to test whether a time lag exists between these
energy bands, we compute the cross-correlation function between the
three energy bands using the {\sc xronos} package (Stella \& Angelini
1991) as well as the discrete correlation function (DCF; Edelson \&
Krolik 1988).  We measured the hard-to-soft band lag to be $230 \pm 50
\s$ (i.e., the hard band lags the soft band by this amount of time;
Fig.~11).

\begin{figure}
\includegraphics[scale=.5,angle=270]{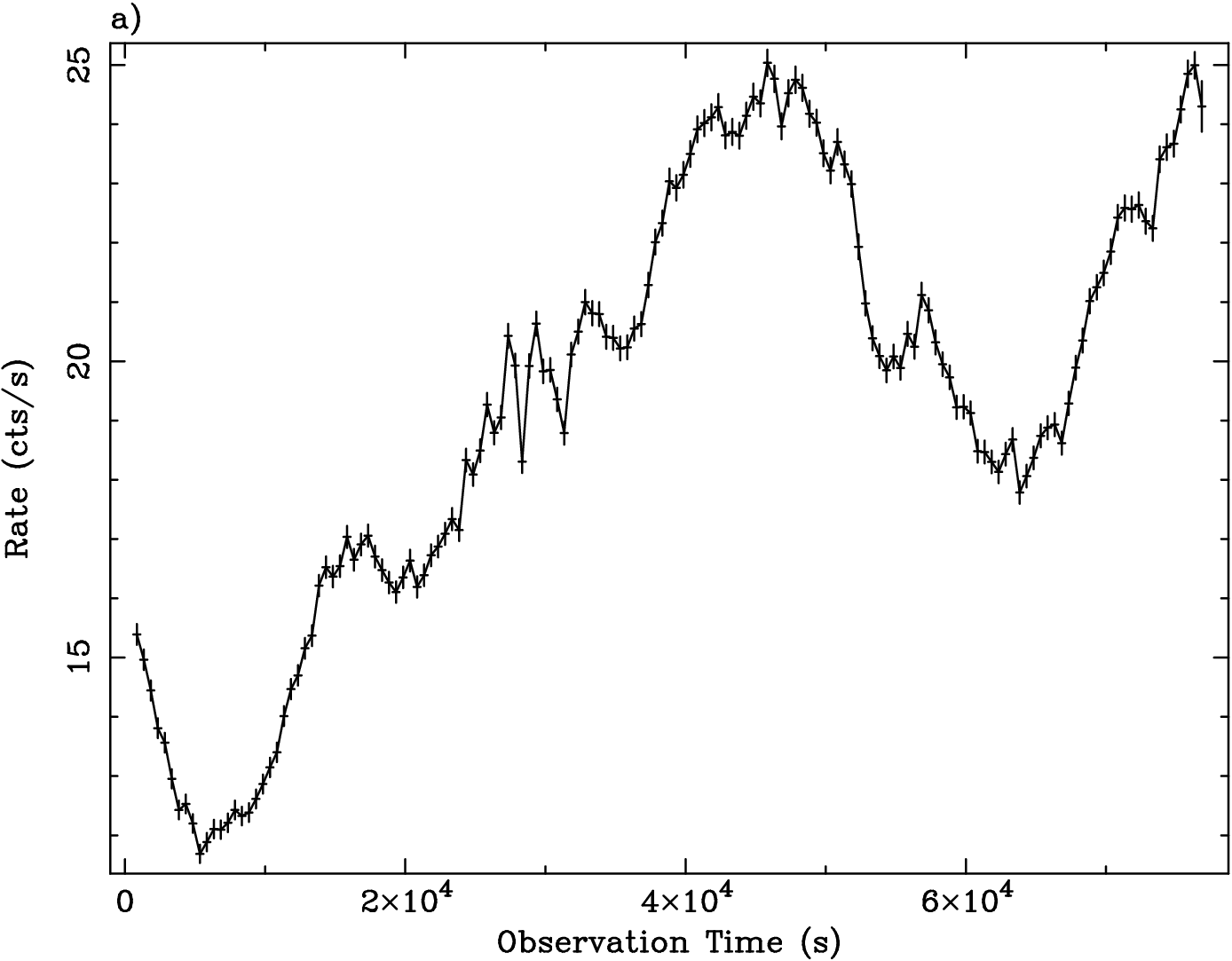}\\
\includegraphics[scale=.5,angle=270]{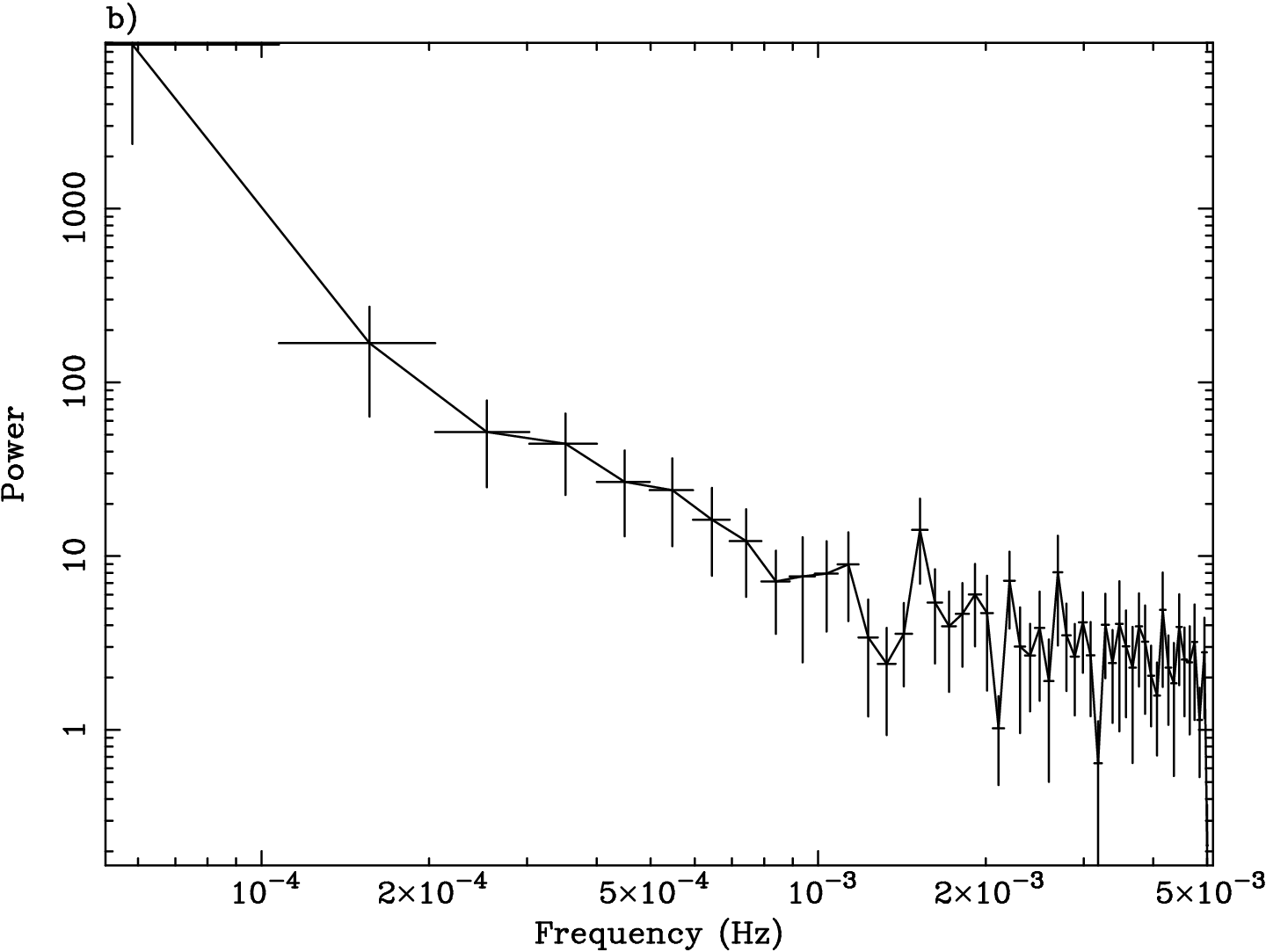}
\caption{The light curve (a) and power spectrum (b) for the
time-averaged pn data from NGC~4593.  Variations on timescales as small
as hundreds of seconds appear visible in the source.  Note that the
frequency at which the power spectrum flattens into Poisson noise is
about $5 \times 10^{-3} \Hz$.  Inverting this means that the smallest
timescale of variability we can reliably observe from this source is
$\sim 200 \s$.  At low frequencies, the slope of the power spectral
density curve is $-2.63^{+0.66}_{-0.25}$.}
\end{figure}

\begin{figure}
\centerline{
\includegraphics[scale=.7,angle=270]{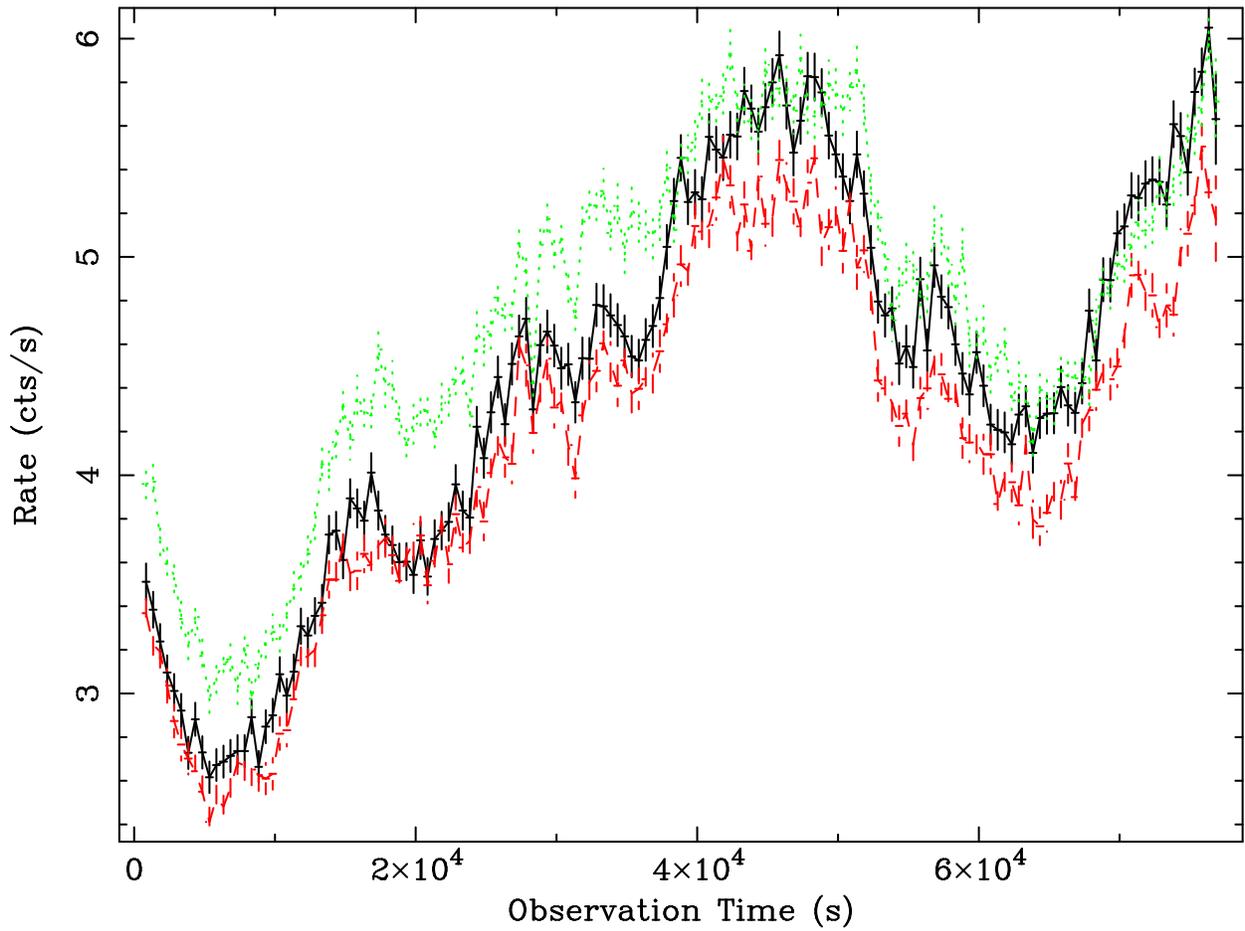}
}
\caption{Light curves for the different energy bands: the soft band
runs from $0.5-0.84 \keV$ (solid black), the mid band runs from $0.84-1.5
\keV$ (dashed red), and the hard band runs from $1.5-10.0 \keV$
(dotted green).}
\end{figure}

The most straightforward interpretation of these time delays envisions
them as representing the 
finite scattering time within the Comptonizing
disk-corona system that is thought to be responsible for the primary X-ray
production.  One should be cautioned, however, that this
interpretation invokes numerous underlying simplifications with regard
to the physical mechanism that produces these types of time lags between
energy bands.  See Pottschmidt \etal (2003) and Nowak \etal (1999,
2002) for more detailed accounts of other factors that can be
responsible for or affect the observed time lag between the hard and
soft energy bands.  We assume here that both the hard and soft X-ray
photons originate from seed UV photons from the disk, which are then
Comptonized by a corona of relativistic electrons surrounding the
disk.  The energy of a given photon in our simplified scheme will
thus depend on the number of scatterings it undergoes with the
electrons, and one may therefore infer that the more energetic photons
have experienced more interactions.  It is thus possible to estimate
the size of the coronal region by measuring the time lag between the
peaks of the hard and soft photon light curves of the source, as we
have done, and making the simplified assumption that the corona
possesses a slab or spherical geometry.  With each scattering, a seed
photon gains on average a fractional energy $\Delta E/E \approx
4k_BT/m_ec^2$, where $T$ is the temperature of the electrons in the
corona.  So the energy of a Comptonized photon after a given number of
scatterings 
will be:
\begin{equation}
E \approx E_0 \left(1 + \frac{4k_BT}{m_ec^2}\right)^n
\end{equation}
where $E_0$ is the initial seed photon energy and $n$ is the number of
scatterings (Rybicki \& Lightman 1979; Chiang et al. 2000).  The time
delay of this photon relative to the seed photon source will be
roughly proportional to the number of scatterings, $t \approx nt_0$.
Here $t_0 \sim l_T/c$ where $l_T$ is the mean free path for Thomson
scattering in an optically-thick corona or, if the corona is
optically-thin, simply the size of the corona.  We take the effective 
photon energy in the soft
range ($0.5-0.84 \keV$) to be $0.67 \keV$.  For the hard range
($1.5-10 \keV$) we calculate the effective photon energy to be $2.17
\keV$.  Both estimates are based on taking weighted averages (based on
flux) of the
energies in the soft and hard ranges, and accounting for the
energy-dependent effective area of the pn.  Assuming that the energy of the coronal
electrons ranges from $50-100 \keV$, our measured soft-hard
time lag of $\sim 226 \s$ suggests that the corona occupies a region around
the disk $1.91-3.34 \times 10^{12} \cm$ in size.  This is derived assuming
the size of the corona is given by
\begin{equation}
r \approx tc  \frac{\log \left(1 + \frac{4k_{\rm B} T}{m_{\rm e} c^2}
  \right)}{\log \frac{E}{E_{\rm 0}}}
\end{equation}
We will assume that the central black hole in NGC~4593 has a mass of
$8.1 \times 10^6 \Msun$ based on reverberation mapping (Nelson \&
Whittle 1995; Gebhardt et al. 2000).  Using this mass for the NGC~4593 
black hole, the characteristic length scale of the corona would
be in the range of $1.6-2.8 r_{\rm g}$.

\begin{figure}
\centerline{
\includegraphics[scale=.7,angle=270]{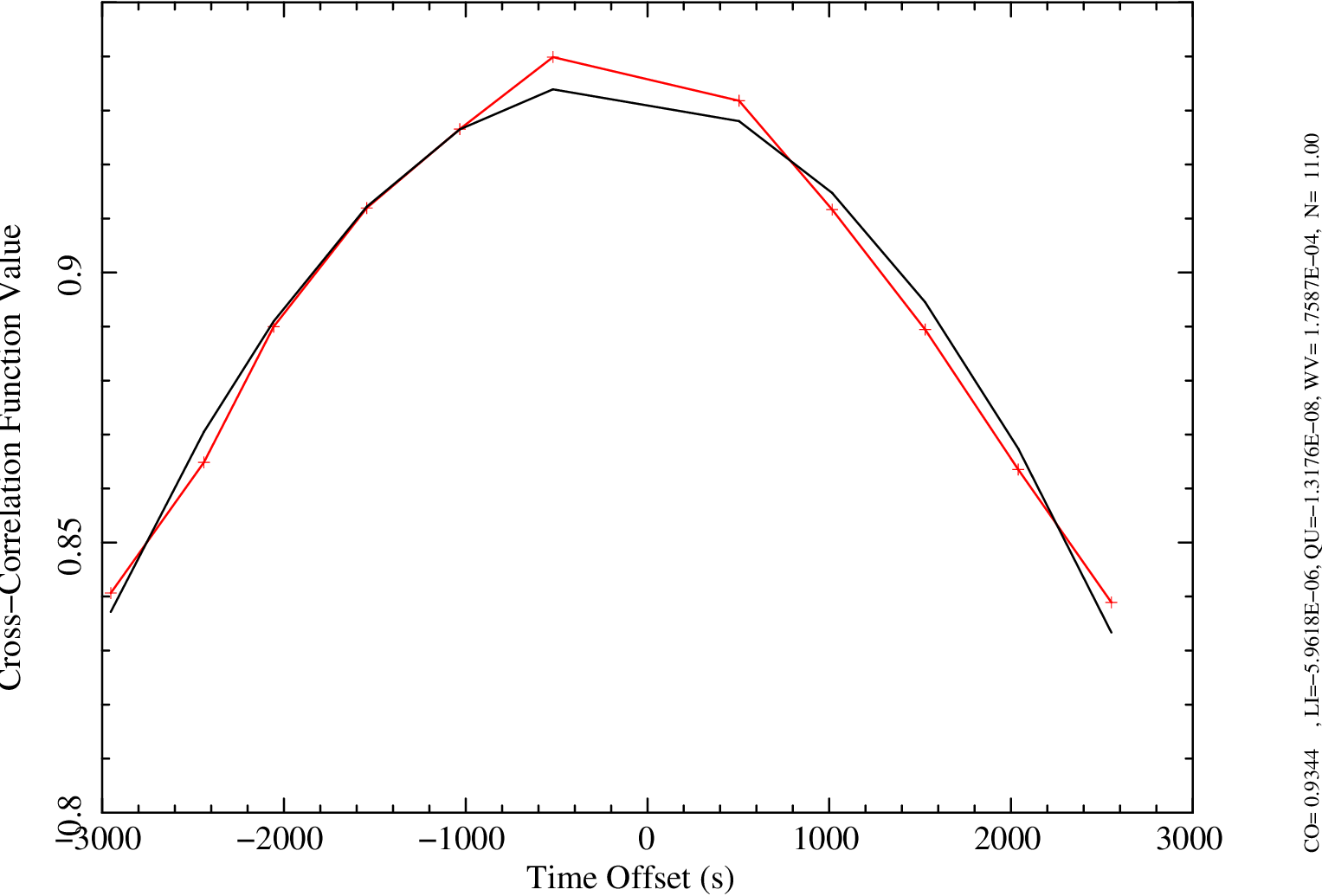}
}
\caption{The cross-correlation function for the soft-to-hard energy
  bands in NGC~4593 in red (with data points), plotted with the best
  fit curve in solid black.  The vertical line represents a zero time
  delay between the two bands; the off-centeredness of the curve peak
  indicates that the hard band lags the soft band by $226 \pm 53 \s$
  at $90\%$ confidence.}
\end{figure}

\section{Discussion and Conclusions}
\label{sec:disc}

\subsection{Summary of Results}
\label{sec:summ}

We have obtained a $76 \ks$ {\it XMM-Newton} observation of the Sy-1
galaxy NGC~4593.  An examination of the best fitting EPIC-pn spectrum
shows that the continuum is well modeled by a photoabsorbed power-law
with $N_{\rm H} = 1.97 \times 10^{20} \pcmsq$, $\Gamma = 1.75$, and a
flux of $4.44 \times 10^{-13} \ergpcmsqps$.  The best fit to the hard
spectrum can be achieved by including two Gaussian emission lines
representing cold and ionized fluorescent iron at $6.4$ and $6.97
\keV$, respectively.

We see clear evidence for a complex warm absorber in the EPIC-pn
spectrum.  Fitting a grid of photoionization models computed using the
{\sc xstar} code, we infer a warm absorber with two physically and
kinematically distinct zones: one with a column density of $N_{\rm
H}=9.29 \times 10^{22}\pcmsq$ and an ionization parameter of
$\log\xi=2.75$, the other with $N_{\rm H}=1.13 \times 10^{22} \pcmsq$
and $\log\xi=1.70$, which is likely more distant from the central
engine.  We robustly detect the L$_3$ edge of neutral iron presumed to
exist in the form of dust grains along the line-of-sight to the
central engine, as cited by McKernan \etal (2003).  This edge has a
column density of $N_{\rm Fe}=8.32 \times 10^{16} \pcmsq$.

A soft excess below $\sim 2 \keV$ is also seen in the data, and can be
accounted for either phenomenalogically by a redshifted bremsstrahlung
component (Model~1) or, more effectively and physically, a component
of Comptonized emission from an accretion disk of seed photons
upscattered by a plasma of relativistic electrons existing in either a
corona or the base of a jet near the disk in some geometry (Model~2).
The latter scenario yields the best statistical fit to the data, with
a seed photon temperature of $T_{\rm 0}=50 \eV$, an electron
temperature of $kT=42 \keV$, a plasma optical depth of $\tau_{\rm
p}=0.12$ and a flux of $6.46 \times 10^{-14} \ergpcmsqps$.

Lastly, cold and ionized Fe-K emission lines were wanted by the fit at
$6.4$ and $6.97 \keV$, respectively.  The cold line Gaussian had a
time-averaged equivalent width of $EW=131 \eV$, while the ionized
component had an $EW=45 \eV$.  Although the continuum varied with time
over the course of our observation, the flux of the cold Fe-K$\alpha$
showed marginal evidence for variability between successive $10 \ks$
intervals.  The equivalent width of this line, on the other hand, is
shown to vary significantly.  The simplest interpretation of this
result is to suppose that the cold line originates from a region with
a light-crossing time larger that the length of our observation.  For
a black hole mass of $8.1 \times 10^6 \Msun$, this places the cold
line emitting region beyond about $2000 r_g$ from the black hole,
i.e., in the outer accretion disk or the putative molecular torus of
Seyfert unification schemes.  Our statistics on the ionized Fe-K line
are insufficient to constrain its variability properties --- these
data are consistent with both constant flux and constant equivalent
width.

We have detected a $226 \s$ time-lag between the hard and soft EPIC-pn
bands, with the hard band lagging the soft.  In a simple model in
which this corresponds to scattering times within a Comptonizing
corona, we conclude that the corona can only possess a size of $\sim
1.6-2.8 r_{\rm g}$.

\subsection{Implications for the X-ray Emission Region}

The narrowness of the fluorescent iron line together with the lack of
response of this line to changes in the hard X-ray continuum suggest
an absence of a cold, optically-thick matter within the central $\sim
10^3 r_g$ of the accretion disk.  The standard framework for
accommodating such a result is to postulate that the inner regions ($r
\approxlt 10^3 r_{\rm g}$) of the accretion flow have entered into a
radiatively-inefficient mode which is extremely hot (electron
temperatures of $T \sim 10^{10} \K$), optically-thin, and
geometrically-thick (e.g., Rees 1982; Narayan \& Yi 1994).  Within
this framework, the hard X-ray source is identified as thermal
bremsstrahlung or Comptonization from this structure.  However, the
variability of the X-ray continuum is inconsistent with X-ray emission
from a structure which is $\sim 1000 r_{\rm g}$ in extent.  The rapid
continuum variability and the short time lag between the hard and soft
X-ray photons dictate that, at any given time, the X-ray emission
region is only $\sim 1-3 r_{\rm g}$ in extent.

We are therefore led to consider alternative geometries and/or origins
for the X-ray source.  We consider three possibilities.  Firstly, the
inner accretion flow may indeed be radiatively-inefficient, but the 
observed X-ray emission may come
from the compact base of a relativistic jet powered by black hole
spin.  Secondly, the X-ray emission may indeed originate from the body
of the radiately-inefficient flow but, at any given instant in time,
be dominated by very compact emission regions within the flow.  Such
emission regions may be arise naturally from the turbulent flow or,
instead, may be related to a magnetic interaction between the inner
parts of the flow and the central spinning black hole (Wilms et
al. 2001; Ye et al. 2007).  In this case the iron line could still
plausibly originate at a distance of thousands of $r_{\rm g}$ and thus
be quite narrow.  Finally, it is possible that the accretion disk is
radiatively-{\it efficient} (and hence optically-thick and
geometrically-thin) close to the black hole and supports a compact
accretion disk corona.  Of course, an immediate objection to this
scenario is the lack of a broadened iron line.  However, as
demonstrated in R04, it is possible the iron line is so broad that it
is buried in the noise of the continuum.  Alternatively, the accretion
disk surface may have an ionization state such that iron line photons
are effectively trapped by resonant scattering and destroyed by the
Auger effect.  Longer observations with better spectral and timing
resolution will be necessary in order to confirm our results and
differentiate between these different possible scenarios.

\subsection{Conclusions}
\label{sec:conclusions}

We have shown that the Sy-1 galaxy NGC~4593 has a continuum spectrum
that is fit remarkably well by a simple photoabsorbed power-law above
$\sim 2 \keV$.  Below this energy, we see evidence for spectral
complexity that can be attributed to the presence of a possible
multi-zone layer of absorbing material intrinsic to the source, as
well as a soft excess that cannot be explained by a reflection model
from an ionized disk.  Also arguing against a
disk-reflection-dominated source is that
unlike other sources of its kind (e.g., MCG--6-30-15), NGC~4593 has
relatively narrow cold and ionized Fe-K$\alpha$ line features at $6.4$
and $6.97 \keV$, respectively.  See R04 for a more 
thorough discussion of these lines and their
possible physical interpretations.  We can say that the cold line is
most likely formed either quite far out in the accretion disk, or
possibly in the putative dusty torus region surrounding the central
engine, based on their narrowness and lack of significant variability
over the $76 \ks$ duration of our observation.  We find no evidence
for reflection features from the inner accretion disk (e.g., a ``broad
iron line'') in the spectrum.

Based on the time lag between the soft and hard spectral bands, we
estimate that the corona occupies the a region around the central
source on the order of $\sim 1.6-2.8 r_{\rm g}$, assuming that
NGC~4593 harbors a $8.1 \times 10^6 \Msun$ black hole at its core and that
the energies of the electrons in the corona range from $\sim 50-100
\keV$.  This estimate for the coronal size is reinforced by our
measurements of continuum variability on timescales as small as $\sim
200 \s$, equivalent to a light travel time of $\sim 4 r_{\rm g}$ for
the black hole mass in question.

Taken together, the implications of a narrow iron line emitted far out
in the disk or torus and the small coronal size in NGC~4593 present an
atypical picture of an AGN.  We postulate that the primary X-ray
source is associated with the compact base of a jet or compact
emission regions within a much larger optically-thin accretion
flow.  Alternatively, the accretion disk may be radiatively-efficient
with a compact corona but not display a broad iron line due to the
effects of extreme broadening or disk ionization.

\section*{Acknowledgments}

We thank Andrew Young for numerous helpful conversations throughout
the course of this work.  Julia Lee provided several {\sc xstar} models we
used in attempting to parameterize the spectrum from $0.3-2 \keV$.  Barry
McKernan's expertise on the spectrum of NGC~4593 as seen by {\it
  Chandra} has also proven invaluable.  We acknowledge support from
the NASA/{\it XMM-Newton} Guest Observer Program under grant NAG5-10083, and the
National Science Foundation under grant AST0205990.

\end{document}